\documentclass[aps,pre,preprint,groupedaddress,notitlepage,amsmath,amssymb]{revtex4-2}
\usepackage{bm}
\usepackage{graphicx}
\usepackage{placeins}
\usepackage{hyperref}
\makeatletter
\let\preprint@abstract@produce\frontmatter@abstract@produce
\def\frontmatter@abstract@produce{%
  \begingroup
  \@booleanfalse\preprintsty@sw
  \preprint@abstract@produce
  \endgroup
}
\makeatother

\hypersetup{
  pdftitle={Dynamical mean-field theory for associative memory with block-correlated patterns},
  pdfauthor={Andrea Callegaris, Yoshiyuki Kabashima},
  hidelinks
}

\begin{document}

\title{Dynamical mean-field theory for associative memory with block-correlated patterns}

\author{Andrea Callegaris}
\email{a.callegaris@campus.unimib.it}
\affiliation{
  Department of Physics ``Giuseppe Occhialini'',
  Universit\`a degli Studi di Milano-Bicocca,
  Piazza della Scienza 3, 20126 Milano, Italy
}

\author{Yoshiyuki Kabashima}
\email{kaba@phys.s.u-tokyo.ac.jp}
\affiliation{
  Institute for Physics of Intelligence,
  The University of Tokyo,
  7-3-1 Hongo, Tokyo 113-0033, Japan\\
  Trans-Scale Quantum Science Institute,
  The University of Tokyo,
  7-3-1 Hongo, Tokyo 113-0033, Japan
}

\date{October 29, 2026}

\begin{abstract}
In associative memories storing correlated patterns, retrieval of individual
memories can compete with symmetric mixed states.  A dynamical description is
needed to determine how the initial state selects between these outcomes.
We extend the generating-functional dynamical mean-field theory (DMFT) for
independent patterns to independent blocks of correlated binary patterns of
fixed finite size.  The resulting effective single-neuron process has a
self-consistent cross-talk noise covariance and self-feedback coefficients
that depend on the block-covariance eigenvalues.  For hierarchically correlated
patterns with nonmonotonic transfer, increasing the loading can permit
retrieval from an imperfect initial state before further loading prevents it.
DMFT reproduces the retrieval trajectories and basin geometry found in direct
simulations.  Near the lower boundary of this re-entrant region, direct
networks reveal stationary saddles separating coexisting mixed and
target-retrieval states.  Parameter sensitivities of DMFT trajectories predict
nearby shifts of this boundary.  At the tested upper boundary, retrieval
remains accessible from initial states with larger target overlap, indicating
a change in basin access.  Under fixed-point convergence assumptions, the
theory recovers the equations of generalized self-consistent signal-to-noise
analysis.
\end{abstract}

\maketitle

\section{Introduction}

Associative memory models describe how recurrent neural networks retrieve
stored patterns from corrupted initial states.  In the Hopfield model, random
binary patterns are embedded in the synaptic couplings by Hebb's rule
\cite{Hopfield1982}.  The statistical mechanics of this model is well understood
when the patterns are independent \cite{AmitEtAl1985PRL,AmitEtAl1987}.

Memories with common structure require correlated patterns.  Hierarchical and
ultrametric ensembles have therefore been studied since the early development
of associative memory
\cite{PargaVirasoro1986,Gutfreund1988,BacciMatoParga1990,Tsodyks1990}.
These models support retrieval at different levels of a hierarchy; in
particular, Gutfreund described retrieval of a descendant preceded and
assisted by retrieval of its ancestors \cite{Gutfreund1988}.
Correlated-pattern models have also been studied in relation to storage
capacity \cite{FontanariTheumann1990} and symmetric and asymmetric mixed states,
which overlap with several stored patterns \cite{ErichsenTheumann1991}.

Ref.~\cite{ToyaEtAl2000} studied independent clusters whose descendants are
generated from a latent ancestor and whose descendants alone enter the coupling.
The generalized self-consistent signal-to-noise analysis (SCSNA) describes the
stationary states and predicts two coexisting symmetric mixed states, each with
equal overlaps with the descendants of a cluster.  Simulations show that
retrieval can approach a mixed state before selecting one memory pattern.

The remaining question concerns the dynamics that selects among these states.
A stationary analysis identifies possible endpoints but does not determine the
finite-time trajectory or its dependence on the initial state.  Retrieval can
also occur transiently when stable retrieval states are absent \cite{Clark2026}.
This distinction is especially important for nonmonotonic transfer functions.
Such functions can increase storage capacity and alter the attractor structure
\cite{Morita1993,YoshizawaEtAl1993,FukaiKimShiino1995}.  The resulting dynamics
does not generally follow an energy function.

Ref.~\cite{KabashimaMimura2026} analyzed this dynamics using the dynamical mean
field theory (DMFT) for independent patterns.  The analysis relates the basin
of attraction to the cross-talk noise and the self-feedback coefficients.  It
also shows how the finite-time trajectory selects a branch when the stationary
fixed-point condition has several roots.  This independent-pattern theory
provides the starting point for the present work.

We extend this analysis to correlated patterns in independent blocks of
fixed finite size.  The block-covariance eigenvalues enter both the cross-talk noise
covariance and the self-feedback coefficients.  We apply it to the
hierarchically correlated patterns of Ref.~\cite{ToyaEtAl2000}.

With nonmonotonic transfer, increasing the loading can restore retrieval from
an imperfect initial state before retrieval fails again at higher loading.
Near the lower boundary, direct simulations identify competition between
coexisting mixed and target-retrieval states.  Parameter sensitivities of the
DMFT trajectories predict nearby shifts of this boundary.

This paper is organized as follows.  In the next section, we set up the model
we analyze.  Section~\ref{sec:dmft} derives the DMFT equations and describes
their numerical solution.  Section~\ref{sec:results} compares the theory with
direct simulations and examines retrieval dynamics and basins of attraction.
Section~\ref{sec:summary} summarizes the results and their implications.
Appendices~\ref{app:derivation} and \ref{app:fixed} give the generating-functional
derivation and the stationary reduction to generalized SCSNA, respectively.

\section{Block-correlated associative memory}
\label{sec:model}

We consider $N$ neurons and $p$ independent blocks.  Each block contains
$s$ correlated binary memory patterns.  The block size $s$ remains fixed to
$O(1)$ as $N\to\infty$.  At site $i$, block $\mu$ is represented by
\begin{equation}
 \bm\xi_i^\mu=(\xi_i^{\mu,1},\ldots,\xi_i^{\mu,s})
 \in\{-1,+1\}^s.
 \label{eq:block}
\end{equation}
The vectors are independent over $i$ and $\mu$ and have distribution
$P(\bm\xi)$.  We use $\mathbb{E}_X[\cdot]$ for expectation over $X$ and assume
\begin{equation}
 \mathbb{E}_{\bm\xi}[\xi_\nu]=0,
 \qquad
 B_{\nu\nu'}=\mathbb{E}_{\bm\xi}[\xi_\nu\xi_{\nu'}],
 \qquad B_{\nu\nu}=1.
 \label{eq:B}
\end{equation}
The loading rate $\alpha=p/N$ is held fixed in the thermodynamic limit.
The number of stored patterns per neuron is $s\alpha$.

The memory patterns determine the Hebbian coupling,
\begin{equation}
 J_{ij}=\frac{1-\delta_{ij}}{N}
 \sum_{\mu=1}^{p}\sum_{\nu=1}^{s}
 \xi_i^{\mu,\nu}\xi_j^{\mu,\nu},
 \label{eq:J}
\end{equation}
where $\delta_{ij}$ is the Kronecker delta.  Let $a_i(t)$ be the local
potential, $x_i(t)$ the transfer function output, $\gamma$ the time step
and $\theta_i(t)$ an external field used to define the response.  The
parallel dynamics is
\begin{equation}
 a_i(t+1)=a_i(t)+\gamma\left[-a_i(t)
 +\sum_{j\ne i}J_{ij}x_j(t)+\theta_i(t)\right].
 \label{eq:dynamics}
\end{equation}
The external field $\theta_i(t)$ is set to zero after differentiation.
We use $a_i(0)=0$ and prescribe a binary initial state with
$s_i(0)=x_i(0)=s_i^0$.  For $t\geq1$, the output and binary readout are
\begin{equation}
 x_i(t)=f[a_i(t)],
 \qquad s_i(t)=\operatorname{sign}[a_i(t)].
 \label{eq:outputs}
\end{equation}
We use $\operatorname{sign}(0)=+1$.
The nonmonotonic transfer function is
\begin{equation}
 f(a)=\frac{1-e^{-ca}}{1+e^{-ca}}
 \frac{1+\kappa e^{c'(|a|-h)}}{1+e^{c'(|a|-h)}}.
 \label{eq:transfer}
\end{equation}
Unless stated otherwise, we use
$(c,c',h,\kappa)=(50,15,0.5,-0.5)$ and $\gamma=0.1$, as in the numerical
analysis of Ref.~\cite{KabashimaMimura2026}.

We study retrieval within block $\mu=1$.  Its output and readout overlaps are
\begin{equation}
 m_\nu(t)=\frac1N\sum_i\xi_i^{1,\nu}x_i(t),
 \qquad
 M_\nu(t)=\frac1N\sum_i\xi_i^{1,\nu}s_i(t).
 \label{eq:overlaps}
\end{equation}
This block is called condensed: its overlaps can remain of order unity as
$N\to\infty$, while overlaps with patterns in the other blocks vanish.
The initial state is obtained by independently flipping each bit of the
retrieval pattern $\xi^{1,1}$ with probability $(1-M^0)/2$, so that
\begin{equation}
 P_0(s_i^0\mid\bm\xi_i^1)
 =\frac{1+M^0s_i^0\xi_i^{1,1}}{2},
 \label{eq:initial}
\end{equation}
where $M^0$ is the initial target overlap.  In the thermodynamic limit,
the initial overlaps within the block are therefore
\begin{equation}
 m_\nu(0)=M_\nu(0)=M^0B_{\nu1}.
 \label{eq:initial-overlaps}
\end{equation}

The output correlation and response function are
\begin{align}
 Q(t,t')&=\frac1N\sum_i\mathbb{E}_{\bm s^0}[x_i(t)x_i(t')],
 \label{eq:Qdef}\\
 G(t,t')&=\frac1N\sum_i
 \left.\frac{\partial\mathbb{E}_{\bm s^0}[x_i(t)]}
 {\partial\theta_i(t')}\right|_{\bm\theta=0}.
 \label{eq:Gdef}
\end{align}
These expectations average over the initial state
$\bm s^0=(s_1^0,\ldots,s_N^0)$ at fixed stored patterns.
The external field at $t'$ first affects $a(t'+1)$.  Hence $G(t,t')=0$ for
$t\leq t'$.

Our main example is the ensemble of hierarchically correlated patterns in
Ref.~\cite{ToyaEtAl2000}.  Each block $\mu$ is generated from an ancestor
pattern $\zeta^\mu$ whose entries independently take the values $\pm1$ with
equal probability.  For $0\leq b\leq1$, the descendant distribution at each
site is
\begin{equation}
 P_b(\bm\xi)=\frac12\sum_{\zeta=\pm1}
 \prod_{\nu=1}^{s}\frac{1+b\zeta\xi_\nu}{2}.
 \label{eq:Pb}
\end{equation}
Thus $\mathbb{E}_{\zeta,\bm\xi}[\zeta\xi_\nu]=b$ and
$\mathbb{E}_{\bm\xi}[\xi_\nu\xi_{\nu'}]=b^2$ for $\nu\ne\nu'$.  The ancestor is
not stored.  With $I_s$ denoting the $s\times s$ identity matrix and $\bm 1$
the $s$-component vector of ones, the covariance is
\begin{equation}
 B=(1-b^2)I_s+b^2\bm 1\bm 1^{\mathsf T}.
 \label{eq:Bexchange}
\end{equation}
Its eigenvalues are \cite{ToyaEtAl2000}
\begin{equation}
 \lambda_+=1+(s-1)b^2,
 \qquad
 \lambda_-=1-b^2,
 \label{eq:lambdapm}
\end{equation}
with multiplicities one and $s-1$.  The eigenvalue
$\lambda_+$ belongs to the symmetric block direction $\bm 1$, in which all
descendants vary together.  The eigenvalue $\lambda_-$ belongs to the $s-1$
directions orthogonal to $\bm 1$, which distinguish one descendant
from another.  At $b=0$, both
eigenvalues equal one.  The descendants are then independent patterns with
total loading $s\alpha$.  For $s=1$, the model reduces directly to the
independent-pattern model with loading rate $\alpha$.

As a reference for symmetric mixed states, we use the majority pattern
\begin{equation}
 \eta_i=\operatorname{sign}\!\left(\sum_{\nu=1}^{s}
 \xi_i^{1,\nu}\right),
 \label{eq:eta}
\end{equation}
with $s$ odd so that ties cannot occur.  All calculations involving this
pattern below use $s=3$.  The corresponding readout overlap is
$M_\eta=N^{-1}\sum_i\eta_i s_i$.  We also monitor the ancestor overlap
$M_\zeta=N^{-1}\sum_i\zeta_i^1s_i$.  For $s=3$, the overlap of the majority
pattern with each descendant is $\mathbb{E}_{\bm\xi}[\xi_\nu\eta]=(1+b^2)/2$.
A stationary symmetric mixed state at finite loading may differ from this
majority pattern.

\section{Dynamical mean-field theory}
\label{sec:dmft}

We write the trajectory generating function and separate the recurrent field
into signal and cross-talk terms.  We then diagonalize the covariance of each
uncondensed block and average over its eigenmodes.  The saddle-point action
reduces to an effective single-site process with self-consistent noise and
self-feedback coefficients.
This follows the independent-pattern construction of
Ref.~\cite{KabashimaMimura2026}, with each block-covariance eigenvalue retained
inside the temporal response matrices.

\subsection{Generating function}

We observe outputs at $t=0,\ldots,T$ and integrate the paths through
$a_i(T+1)$, with $a_i(0)=0$ fixed.  The path measure is
\begin{equation}
 D\bm a=\prod_{i=1}^{N}\prod_{t=1}^{T+1}da_i(t).
 \label{eq:pathmeasure}
\end{equation}
Introduce the observable source $\psi_i(t)$, collected in $\bm\psi$, and
write $P_0(\bm s^0)=\prod_iP_0(s_i^0\mid\bm\xi_i^1)$.  The generating function is
\begin{align}
 Z[\bm\psi,\bm\theta]
 ={}&\sum_{\bm s^0}P_0(\bm s^0)
 \int D\bm a\prod_{i=1}^{N}\prod_{t=0}^{T}
 \delta\!\left\{a_i(t+1)-a_i(t)\right.\nonumber\\
 &\left.\quad-\gamma\left[-a_i(t)+\sum_{j\ne i}J_{ij}x_j(t)
 +\theta_i(t)\right]\right\}
 \exp\!\left[i\sum_{i=1}^{N}\sum_{t=0}^{T}\psi_i(t)x_i(t)\right].
 \label{eq:Z}
\end{align}
It satisfies $Z[\bm0,\bm\theta]=1$ for every external field $\bm\theta$.
Its source derivatives give, for $0\leq t,t'\leq T$,
\begin{align}
 \mathbb{E}_{\bm s^0}[x_i(t)]
 &=\left.\frac{\partial Z}{\partial[i\psi_i(t)]}\right|_0,
 \label{eq:Zx}\\
 G(t,t')&=\frac1N\sum_i
 \left.\frac{\partial^2 Z}
 {\partial[i\psi_i(t)]\partial\theta_i(t')}\right|_0,
 \label{eq:ZG}
\end{align}
where $|_0$ sets both $\bm\psi$ and $\bm\theta$ to zero.
Each delta function is written as
\begin{equation}
 \delta(F)=\int_{-\infty}^{\infty}\frac{dh}{2\pi}
 e^{-ih F}.
 \label{eq:Fourier}
\end{equation}
Following Ref.~\cite{KabashimaMimura2026}, we denote the auxiliary
Fourier field by $h_i(t)$.  The Fourier variables expose
the pattern dependence in bilinear form, allowing the trajectories to be
averaged over the stored patterns.

The recurrent field separates into a condensed signal, uncondensed
cross-talk noise and the exact diagonal subtraction,
\begin{align}
 \sum_{j\ne i}J_{ij}x_j(t)
 ={}&\sum_{\nu=1}^{s}\xi_i^{1,\nu}m_\nu(t)
 +\frac1{\sqrt N}\sum_{\mu=2}^{p}\sum_{\nu=1}^{s}
 \xi_i^{\mu,\nu}u_{\mu\nu}(t)
 -\frac{ps}{N}x_i(t),
 \label{eq:fieldsplit}\\
 u_{\mu\nu}(t)&=\frac1{\sqrt N}\sum_j
 \xi_j^{\mu,\nu}x_j(t).
 \label{eq:u}
\end{align}
For the Fourier variables we also define
\begin{equation}
 v_{\mu\nu}(t)=\frac1{\sqrt N}\sum_i
 \xi_i^{\mu,\nu}i h_i(t).
 \label{eq:v}
\end{equation}
Substituting into Eq.~\eqref{eq:Z}, the uncondensed blocks contribute the
disorder average
\begin{equation}
 \prod_{\mu=2}^{p}\mathbb{E}_{\{\bm\xi_i^\mu\}}
 \exp\left[\gamma\sum_{\nu=1}^{s}\sum_{t=0}^{T}
 v_{\mu\nu}(t)u_{\mu\nu}(t)\right].
 \label{eq:disorderaverage}
\end{equation}
To evaluate this average, we hold the paths and Fourier variables fixed
and introduce the empirical temporal matrices
\begin{align}
 Q(t,t')&=\frac1N\sum_i x_i(t)x_i(t'),\nonumber\\
 R(t,t')&=\frac1N\sum_i[i h_i(t)][i h_i(t')],
 \label{eq:QRS}\\
 S(t,t')&=\frac1N\sum_i x_i(t)i h_i(t').\nonumber
\end{align}
The amplitudes have zero mean and second moments
\begin{align}
 \mathbb{E}_{\{\bm\xi_i^\mu\}}[u_{\mu\nu}(t)u_{\mu\nu'}(t')]
 &=B_{\nu\nu'}Q(t,t'),\nonumber\\
 \mathbb{E}_{\{\bm\xi_i^\mu\}}[v_{\mu\nu}(t)v_{\mu\nu'}(t')]
 &=B_{\nu\nu'}R(t,t'),
 \label{eq:tensorcov}\\
 \mathbb{E}_{\{\bm\xi_i^\mu\}}[u_{\mu\nu}(t)v_{\mu\nu'}(t')]
 &=B_{\nu\nu'}S(t,t').\nonumber
\end{align}
At fixed $s$ and $T$, these amplitudes are normalized sums of independent
site contributions.  Their Gaussian limit determines the leading extensive
contribution to the disorder average, as in the independent-pattern
calculation.

Let $W$ be orthogonal and satisfy
\begin{equation}
 WBW^{\mathsf T}=\operatorname{diag}(\lambda_1,\ldots,\lambda_s).
 \label{eq:W}
\end{equation}
Rotating the amplitudes of each block by $W$ preserves the bilinear in
Eq.~\eqref{eq:disorderaverage} and separates the Gaussian
average into modes with second moments $\lambda_kQ$, $\lambda_kR$ and
$\lambda_kS$.  For $\lambda_k>0$, dividing both amplitudes by
$\sqrt{\lambda_k}$ gives the independent-pattern Gaussian integral with
coupling $\gamma\lambda_k$.  Thus Eq.~\eqref{eq:disorderaverage} becomes
\begin{equation}
 \prod_{k=1}^{s}
 \left\{\det Q\,
 \det\!\left[
 (I-\gamma\lambda_kS)^{\mathsf T}Q^{-1}
 (I-\gamma\lambda_kS)-\gamma^2\lambda_k^2R
 \right]\right\}^{-\alpha N/2}
 \label{eq:detproduct}
\end{equation}
to leading order in $N$, where $I$ is the identity on the $T+1$ update
times.  A zero eigenvalue contributes a factor of one.
Each eigenvalue remains inside its temporal determinant.
A related factorization over pattern-space eigenmodes appears in the
generating-functional analysis of finite cycles~\cite{MimuraEtAl2004}.
The definitions of $m_\nu$, $Q$, $R$ and $S$ are imposed by Fourier
constraints, and the remaining integral is evaluated at its saddle point.
At zero sources, normalization and differentiation with respect to the
external field give
\begin{equation}
 R=0,\qquad G=\gamma S.
 \label{eq:GS}
\end{equation}
Appendix~\ref{app:derivation} gives the Gaussian integral and the saddle
equations.

\subsection{Langevin equations}

At the saddle point, the neuronal paths factorize.  The uncondensed blocks
and the diagonal subtraction contribute the following terms to the
single-site exponent:
\begin{equation}
 -\frac{\gamma^2}{2}\sum_{t,t'=0}^{T}h(t)C(t,t')h(t')
 +i\gamma\sum_{t,t'=0}^{T}h(t)\Lambda(t,t')x(t').
 \label{eq:effective-exponent}
\end{equation}
The cross-talk noise covariance and self-feedback coefficients are
\begin{equation}
 \boxed{
 C=\alpha\sum_{k=1}^{s}\lambda_k^2
 (I-\lambda_kG)^{-1}Q
 (I-\lambda_kG^{\mathsf T})^{-1}}
 \label{eq:C}
\end{equation}
and
\begin{equation}
 \boxed{\begin{aligned}
 \Lambda&=\alpha\sum_{k=1}^{s}\lambda_k(I-\lambda_kG)^{-1}
 -\alpha sI\\
 &=\alpha\sum_{k=1}^{s}\lambda_k^2G(I-\lambda_kG)^{-1}.
 \end{aligned}}
 \label{eq:Lambda}
\end{equation}
The instantaneous contribution cancels because
$\sum_k\lambda_k=\operatorname{Tr}B=s$.
Thus $\Lambda(t,t')$ describes self-feedback from earlier outputs.

Introduce a zero-mean Gaussian process $\bm\phi=(\phi(0),\ldots,\phi(T))$
with covariance $C$.  The identity
\begin{equation}
 \exp\left[-\frac{\gamma^2}{2}\sum_{t,t'=0}^{T}h(t)C(t,t')h(t')\right]
 =\mathbb{E}_{\bm\phi}
 \exp\left[i\gamma\sum_{t=0}^{T}h(t)\phi(t)\right]
 \label{eq:Gaussian-decoupling}
\end{equation}
replaces the quadratic Fourier term by the cross-talk noise.
Integration over $h(t)$ then gives the effective Langevin dynamics
\begin{equation}
 \boxed{\begin{gathered}
 \begin{aligned}
 a(t+1)=a(t)+\gamma\Bigg[&-a(t)+\bm\xi^{\mathsf T}\bm m(t)
 +\theta(t)+\phi(t)\\
 &+\sum_{t'<t}\Lambda(t,t')x(t')\Bigg],
 \end{aligned}\\[-2pt]
 a(0)=0,\qquad x(0)=s^0,\qquad
 x(t)=f[a(t)]\quad(1\leq t\leq T).
 \end{gathered}}
 \label{eq:Langevin}
\end{equation}
Here $\bm m(t)=(m_1(t),\ldots,m_s(t))^{\mathsf T}$,
$\bm\xi$ has distribution $P(\bm\xi)$ and $s^0$ is drawn from
$P_0(s^0\mid\bm\xi)$ in Eq.~\eqref{eq:initial}.
The Gaussian noise is independent of $\bm\xi$ and $s^0$.

Averaging over the condensed block, the initial state and the cross-talk
noise gives the self-consistency equations
\begin{align}
 m_\nu(t)&=\mathbb{E}_{\bm\xi,s^0,\bm\phi}[\xi_\nu x(t)],
 \label{eq:mself}\\
 Q(t,t')&=\mathbb{E}_{\bm\xi,s^0,\bm\phi}[x(t)x(t')],
 \label{eq:Qself}\\
 G(t,t')&=\left.\frac{\partial\mathbb{E}_{\bm\xi,s^0,\bm\phi}[x(t)]}
 {\partial\theta(t')}\right|_{\bm\theta=0}.
 \label{eq:Gself}
\end{align}
All averages are evaluated at zero external field.
The local response derivative holds $\bm m$, $C$ and $\Lambda$ fixed.
For given $Q$ and $G$, the uncondensed contribution depends only on the
eigenvalues of $B$, while the condensed averages retain the complete
distribution $P(\bm\xi)$.

Strict causality makes $G$ lower triangular with a zero diagonal.  Hence
each inverse in Eqs.~\eqref{eq:C} and \eqref{eq:Lambda} is the finite sum
\begin{equation}
 (I-\lambda_kG)^{-1}
 =\sum_{n=0}^{T}(\lambda_kG)^n.
 \label{eq:polynomial}
\end{equation}
Each additional response carries a factor $\lambda_k$, so the covariance
eigenvalues weight successive responses differently.
The two inverses in $C$ account for propagation along its two time indices.
For the hierarchically correlated patterns, the kernels contain one
$\lambda_+$ term and $s-1$ identical $\lambda_-$ terms.
For independent patterns, $B=I_s$ and the kernels reduce to those of
Ref.~\cite{KabashimaMimura2026} at total loading $s\alpha$.

For numerical solution, we sample $N_{\mathrm{DMFT}}$ effective trajectories
and replace the expectations by population averages~\cite{EissfellerOpper1992}.
The initial overlaps are given by Eq.~\eqref{eq:initial-overlaps},
while $Q(0,0)=1$ and $G(0,0)=0$ give
\begin{equation}
 C(0,0)=\alpha\sum_{k=1}^{s}\lambda_k^2
 =\alpha\operatorname{Tr}B^2.
 \label{eq:C00}
\end{equation}
At time $t$, the sampled outputs determine $\bm m(t)$ and the new row and
column of $Q$.  After evaluating the response row $G(t,t')$ for $t'<t$,
Eqs.~\eqref{eq:C} and \eqref{eq:Lambda}, restricted to times $0,\ldots,t$,
give the new noise covariances and self-feedback coefficients.
We then draw $\phi(t)$ conditional on the noise history and update
$a(t+1)$ using Eq.~\eqref{eq:Langevin}.
The population size $N_{\mathrm{DMFT}}$ controls sampling error in the
thermodynamic-limit process.

For the smooth transfer, we evaluate the response by differentiating
Eq.~\eqref{eq:Langevin} along each trajectory with the noise and
macroscopic quantities fixed, then averaging the resulting
$\partial x(t)/\partial\theta(t')$~\cite{RoyEtAl2019}.
This response calculation requires $O(N_{\mathrm{DMFT}}T^3)$ operations
over $T$ time steps.
For sign transfer, we instead use Gaussian integration by parts, as in
Ref.~\cite{KabashimaMimura2026}:
\begin{equation}
 G(t,t')=\sum_{u<t}\mathbb{E}_{\bm\xi,s^0,\bm\phi}[x(t)\phi(u)]
 [C_{t-1}^{-1}]_{ut'},\qquad t'<t.
 \label{eq:Grec}
\end{equation}
Here $C_{t-1}$ is the principal submatrix of $C$ on times $0,\ldots,t-1$.
Equation~\eqref{eq:Grec} requires this submatrix to be nonsingular;
Appendix~\ref{app:derivation} discusses the singular case.

\section{Results}
\label{sec:results}

We solved Eq.~\eqref{eq:Langevin} with the kernels in
Eqs.~\eqref{eq:C} and \eqref{eq:Lambda} and the self-consistency equations
\eqref{eq:mself}-\eqref{eq:Gself}.  Direct simulations provide a finite-size
comparison with the thermodynamic-limit dynamics.
Unless stated otherwise, $s=3$, $b=0.61$, $N=16000$ and the observation time
is $T=100$.  The figures use $1.25\times10^5$ effective DMFT trajectories.
Boundary refinements use $10^6$ trajectories and, for direct simulations,
64 disorder realizations with eight initial states per realization.

A large target overlap alone does not distinguish individual retrieval from
a symmetric mixed state.  We therefore count the retrieval pattern
$\xi^{1,1}$ as retrieved when
\begin{equation}
 \frac{M_1(T)-\max_{\nu>1}M_\nu(T)}{1-b^2}\geq\frac12,
 \label{eq:retrieval}
\end{equation}
for $s>1$ and $b<1$.  For $s=1$, the criterion is $M_1(T)\geq1/2$.
Non-target retrieval satisfies the same contrast between the largest and
second-largest descendant overlaps.  A symmetric endpoint is labelled
$\eta$ when its overlaps agree within three pairwise standard errors,
$\sqrt{2(1-b^2)/n}$ for $n$ effective trajectories, and their mean is within
$0.01$ of $(1+b^2)/2$.  The remaining endpoints form the failure class.
These finite-time criteria do not require microscopic convergence.  The
fixed-point analysis below additionally requires the stationary assumptions
in Appendix~\ref{app:fixed}.

\subsection{Basin of attraction}

\begin{figure}[htbp]
 \centering
 \includegraphics[width=0.85\textwidth]{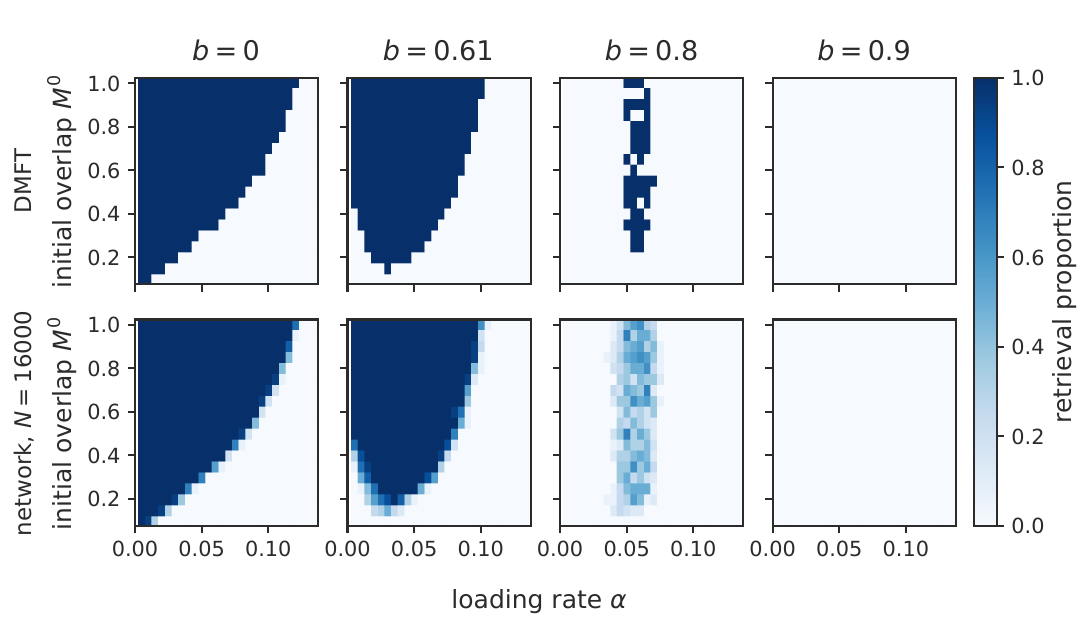}
 \caption{Basin of retrieval pattern $\xi^{1,1}$ according to
 Eq.~\eqref{eq:retrieval}.  Direct-simulation values are retrieval fractions
 over 16 disorder realizations.}
 \label{fig:basin}
\end{figure}

Block correlations change both the retrieval threshold and the dependence
on the initial overlap (Fig.~\ref{fig:basin}).  At $b=0$, the critical total
pattern loading $s\alpha_{\mathrm c}$ agrees with the independent-pattern
result of Ref.~\cite{KabashimaMimura2026}, as required by
Eqs.~\eqref{eq:C} and \eqref{eq:Lambda}.
At finite correlation, DMFT reproduces the basin geometry found in direct
simulations.  Near a boundary, the finite network retrieves in fewer
realizations; in the size comparison, increasing $N$ brings the
finite-network basin closer to the DMFT boundary.

At large $M^0$, the upper boundary moves to lower loading as $b$ increases
from zero to $0.61$.  At small $M^0$, it moves to higher loading over the
same range.  This reversal indicates a competition between the tendency
toward symmetric mixed states and the retrieval of one descendant.

\begin{figure}[htbp]
 \centering
 \includegraphics[width=0.85\textwidth]{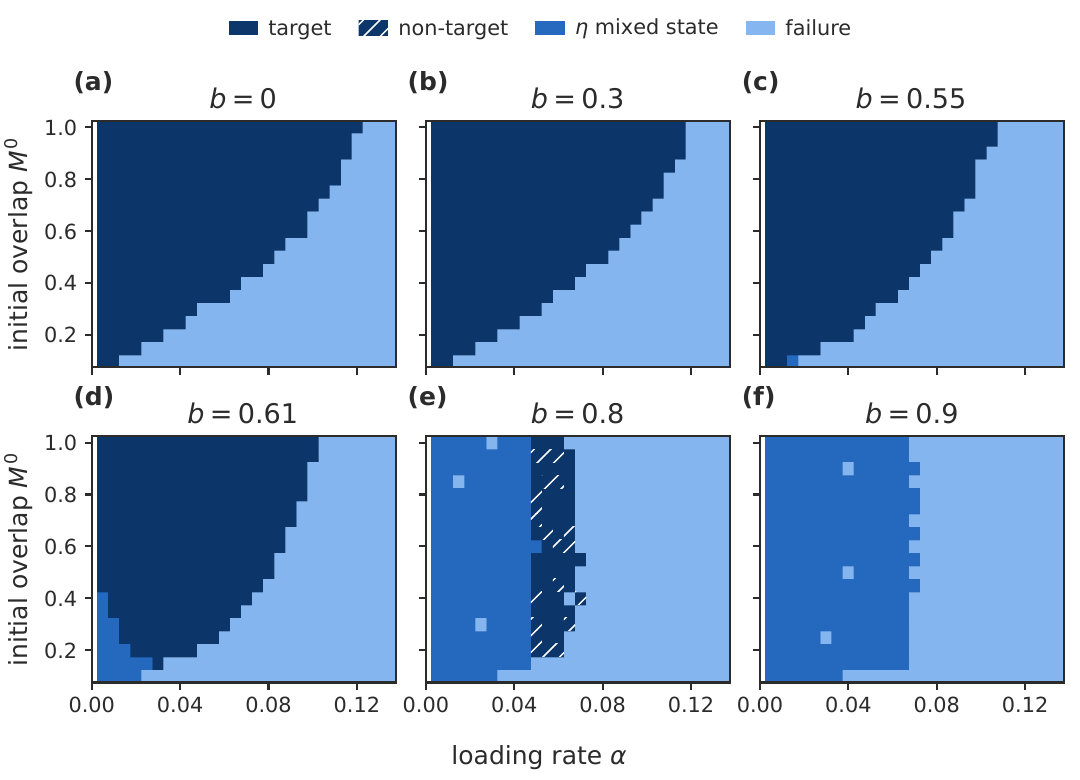}
 \caption{DMFT endpoint classes.  The symmetric class $\eta$ is identified
 by equal descendant overlaps near the majority-pattern value
 $(1+b^2)/2$.}
 \label{fig:classes}
\end{figure}

\begin{figure}[htbp]
 \centering
 \includegraphics[width=0.85\textwidth]{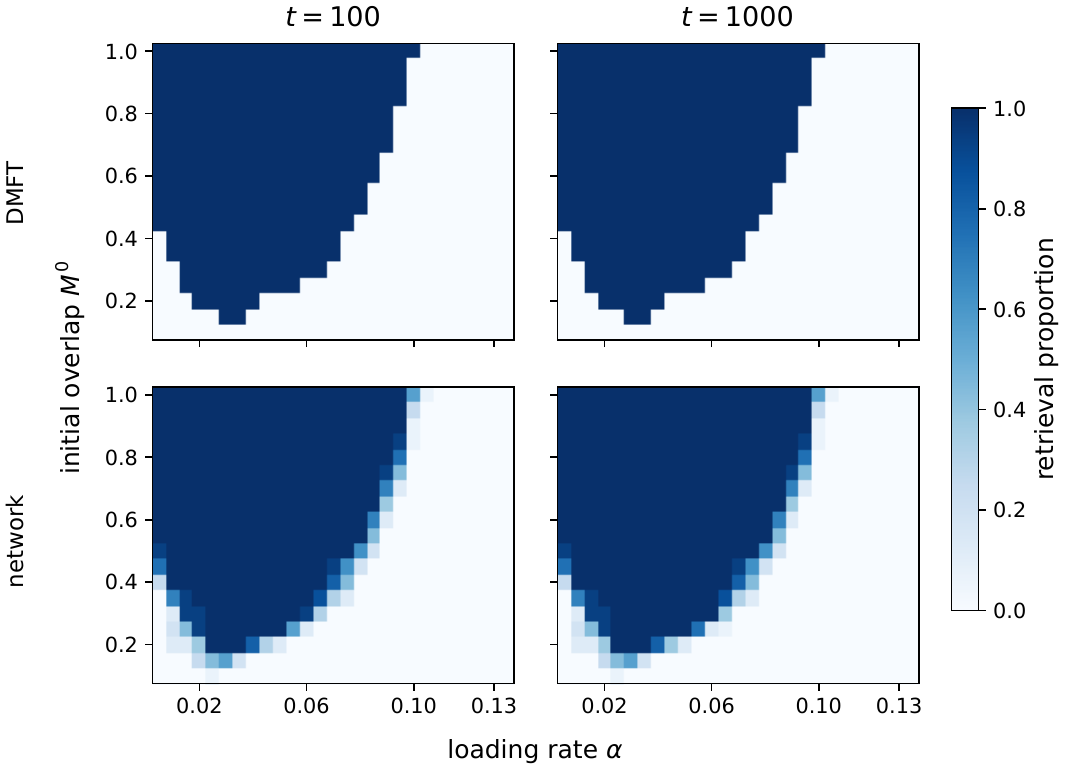}
 \caption{Time dependence of the retrieval basin.  The same trajectories are
 classified at both observation times using Eq.~\eqref{eq:retrieval}.
 Direct-simulation values are retrieval fractions over 16 disorder realizations.}
 \label{fig:basin-time}
\end{figure}

The states that replace target retrieval distinguish these two effects
(Fig.~\ref{fig:classes}).  At $b=0.61$, retrieval from small initial overlaps
occurs only within an intermediate range of loading rates.  Below its lower boundary,
the dynamics reaches the symmetric mixed state.  Increasing the loading
first permits the target to separate from this state.  A further increase
causes the same initial state to lose access to target retrieval.
Figure~\ref{fig:basin-time} compares the retrieval basins at $t=100$ and
$t=1000$ along the same trajectories.  The DMFT classifications are unchanged,
while direct simulations show only small changes near the boundary.
This indicates that the re-entrant geometry reflects converged retrieval
outcomes at the sampled resolution, rather than transient retrieval.
Section~\ref{sec:reentrance} examines the basin selection at both boundaries.

At fixed finite time, exact exchangeability excludes systematic selection of
one non-target descendant in the thermodynamic limit.  The non-target DMFT
outcomes near the boundary at $b=0.8$ become less frequent as the number of
effective trajectories increases at the tested conditions, indicating
symmetry breaking from finite sampling.  The corresponding selection in a
finite network is physical because each disorder realization breaks
descendant exchangeability.

\subsection{Retrieval dynamics and covariance eigenmodes}

At the first update, the effective field has no self-feedback history.
Since $a(0)=0$, expanding the transfer to first order in the condensed
signal gives
\begin{equation}
 \bm m(1)\simeq
 \gamma\,\mathbb{E}_{\phi(0)}
 \bigl[f'(\gamma\phi(0))\bigr]\,B\bm m(0).
 \label{eq:first-update}
\end{equation}
The noise $\phi(0)$ is independent of the condensed pattern vector.
The scalar gain in Eq.~\eqref{eq:first-update} is therefore common to all
directions.  The symmetric output-overlap combination
$s^{-1}\sum_\nu m_\nu$ is amplified in proportion to $\lambda_+$, while
$m_1-(m_2+m_3)/2$, which distinguishes the target from the other descendants,
is amplified in proportion to $\lambda_-$.  Their relative amplification is
\begin{align}
 \mathcal R_1
 &\equiv
 \frac{\bigl[s^{-1}\sum_\nu m_\nu(1)\bigr]
       /\bigl[s^{-1}\sum_\nu m_\nu(0)\bigr]}
      {\bigl[m_1(1)-(m_2(1)+m_3(1))/2\bigr]
       /\bigl[m_1(0)-(m_2(0)+m_3(0))/2\bigr]} \notag\\
 &\simeq \frac{\lambda_+}{\lambda_-}
 =\frac{1+2b^2}{1-b^2}.
 \label{eq:growthratio}
\end{align}
Thus increasing $b$ favours the initial symmetric response relative to
descendant discrimination.  The measured ratios follow this ordering.
Their departure from Eq.~\eqref{eq:growthratio} grows with the initial
overlap, for which the linear expansion is less accurate.  The subsequent
evolution also involves the nonlinear transfer, cross-talk and self-feedback.

\begin{figure}[p]
 \centering
 \includegraphics[width=\textwidth]{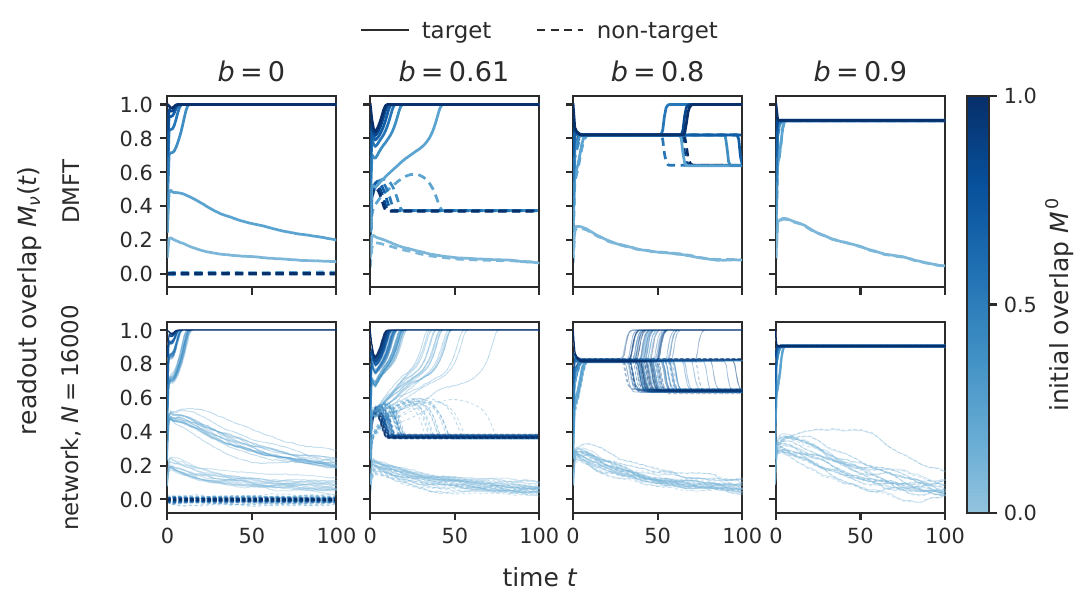}
 \caption{Readout trajectories at $\alpha=0.05$.
 Dashed curves give $(M_2+M_3)/2$.  Direct simulations use 16 disorder
 realizations per initial overlap.}
 \label{fig:readoutcorr}
 \medskip
 \includegraphics[width=\textwidth]{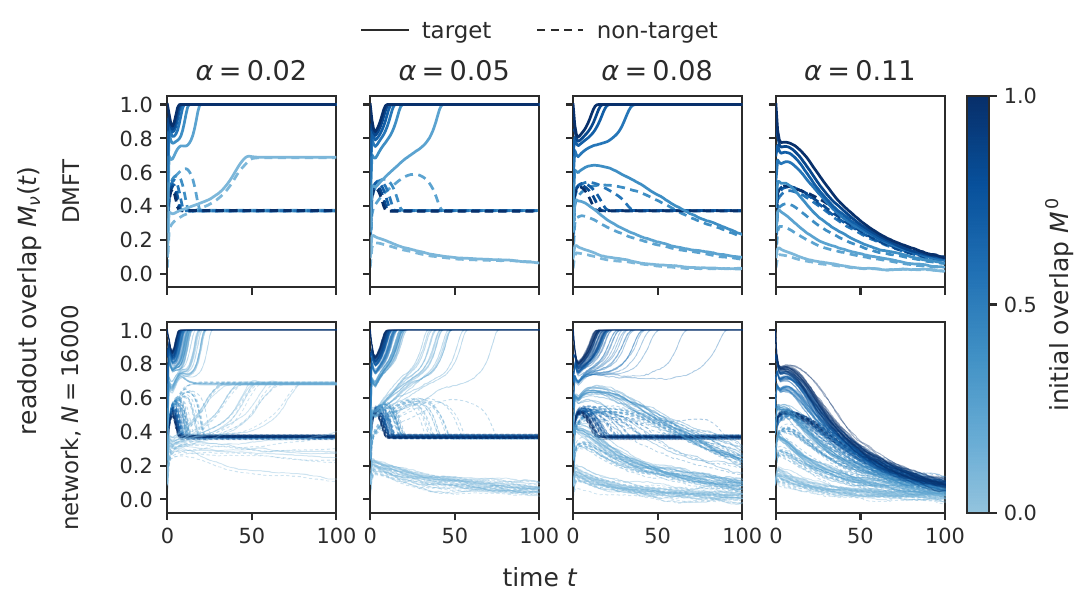}
 \caption{Readout trajectories as the loading increases.
 Dashed curves give $(M_2+M_3)/2$.  Direct simulations use 16 disorder
 realizations per initial overlap.}
 \label{fig:readout}
\end{figure}

The readout trajectories show how this initial tendency develops
(Fig.~\ref{fig:readoutcorr}).  At $b=0.8$ and $M^0=1$,
the descendant overlaps first approach the symmetric mixed-state value and
separate only after a long delay.  At $b=0.9$, the same initial condition
remains at the symmetric state at the observation time, and no target
retrieval is found on the sampled basin grid.  Thus individual retrieval
can fail while the condensed block remains active.

Changing the loading at fixed correlation changes the separation from the
symmetric state (Fig.~\ref{fig:readout}).  On retrieving trajectories, the
non-target descendants increase together before the target separates and
then relax toward $b^2$, their correlation with the retrieved pattern.
These trajectories approach the symmetric mixed state without generally
reaching it.  Near the lower retrieval boundary, they pass closer to the
mixed state; near the upper boundary, target selection becomes slower.
DMFT and direct simulations recover the sequence described in
Ref.~\cite{ToyaEtAl2000}.

The covariance eigenvalues enter two distinct parts of the dynamics.
They act linearly on the initial condensed overlap vector in
Eq.~\eqref{eq:first-update}.  In the uncondensed cross-talk kernels, they
enter exactly as the prefactors $\lambda_k^2$ and inside the temporal
resolvents $(I-\lambda_kG)^{-1}$ in Eqs.~\eqref{eq:C} and \eqref{eq:Lambda}.

\FloatBarrier

\subsection{Basin selection and reentrance}
\label{sec:reentrance}

The lower retrieval boundary marks a change from the symmetric mixed state
to target retrieval.  Both attractors
remain accessible on either side of this boundary in the directly studied
networks.  To identify the separating dynamics, we used edge tracking,
developed to follow the ``edge of chaos'' in shear flows
\cite{SkufcaEtAl2006}.  Potential fields leading to different attractors are
repeatedly bisected and the bracket is advanced with the original dynamics.
At $b=0.61$ and $M^0=0.20$, this procedure finds stationary saddles with one
unstable direction in three networks.  The two unstable branches lead to
mixed and target retrieval.

\begin{figure}[htbp]
 \centering
 \includegraphics[width=\textwidth]{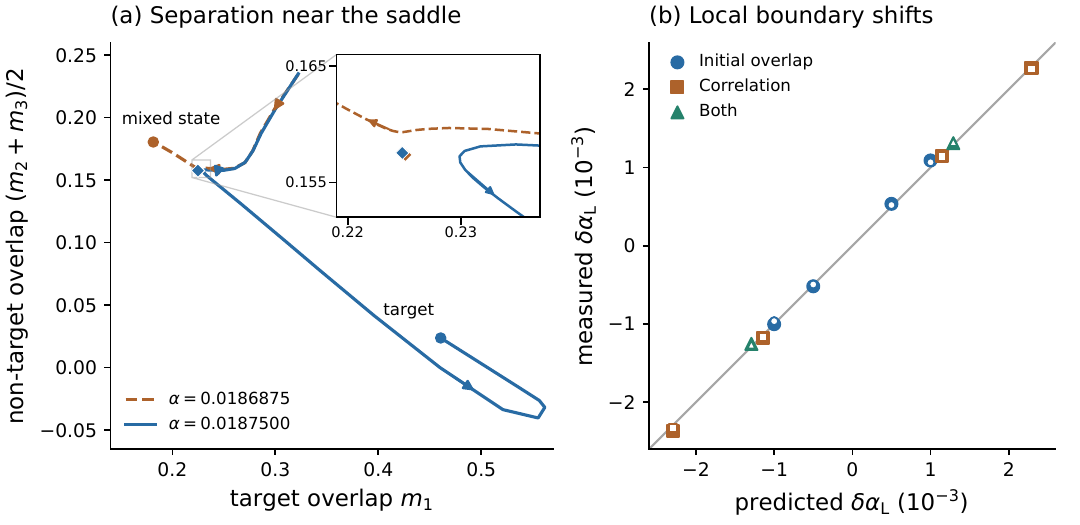}
 \caption{Basin selection near the lower retrieval boundary.
 (a) Direct trajectories from the same initial state with $M^0=0.20$,
 projected onto output overlaps from $t=5$.  The two loadings differ by one
 uncondensed block.  Diamonds mark the saddles and circles the endpoints at
 $T=1000$.  The inset enlarges the saddle neighbourhood.
 (b) Boundary shifts predicted by Eq.~\eqref{eq:boundary-law} and measured
 at ten parameter pairs excluded from coefficient estimation.
 Filled and open markers denote two populations of $10^6$ trajectories.
 Shifts are relative to each population's boundary at $(0.61,0.20)$.}
 \label{fig:reentrance}
\end{figure}

We increased the loading by adding uncondensed blocks while keeping the
condensed patterns and binary initial state fixed in each network.
The resulting trajectories approach different sides of the saddle's stable
manifold (Fig.~\ref{fig:reentrance}a).  At closest approach, the projection of their displacement onto the left
unstable eigenvector identifies the eventual attractor.  Reversing the
corresponding component along the right unstable eigenvector switches the
outcome, whereas equally large
perturbations within the stable eigenspace preserve it.
The lower boundary therefore reflects a change in basin access while both
attractors coexist.  Nearby parameter values also show competition organized
by saddles, although the connections can change: in two of the investigated
cases, a local mixed-state trapping region disappears and a non-target
descendant becomes the competing attractor.

DMFT likewise retains mixed and target retrieval on both sides of the lower
boundary.  Near that boundary, the response of the target-discrimination
overlap,
\begin{equation}
 D(t)=m_1(t)-\frac{m_2(t)+m_3(t)}2,
 \label{eq:discrimination}
\end{equation}
to small perturbations grows, while differences between the two non-target
descendants initially decay.  Moving closer to the boundary delays the
departure from the mixed-state neighbourhood, consistently with an unstable
direction controlling the separation.  The DMFT fields and kernels still
evolve during this passage, so these observations do not establish a
stationary DMFT saddle.

This growing collective response gives a local prediction for the lower
boundary $\alpha_{\mathrm L}(b,M^0)$.  If changes in $\alpha$, $b$ and $M^0$
mainly excite the same growing displacement transverse to the boundary,
their effects cancel for parameter changes along it.  To first order,
\begin{equation}
 \delta\alpha_{\mathrm L}\simeq
 -\frac{\partial_bD(t)}{\partial_\alpha D(t)}\,\delta b
 -\frac{\partial_{M^0}D(t)}{\partial_\alpha D(t)}\,\delta M^0.
 \label{eq:boundary-sensitivity}
\end{equation}
The ratios become approximately time independent when the common growing
response dominates.  These derivatives include the change of the complete
self-consistent DMFT history; $G$ instead describes the response to a local
external field at fixed macroscopic kernels.

Central differences with shared Gaussian variables, averaged over the interval
where the ratios have settled, give at $(b,M^0)=(0.61,0.20)$
\begin{equation}
 \delta\alpha_{\mathrm L}\simeq
 0.4583\,\delta b-0.1000\,\delta M^0.
 \label{eq:boundary-law}
\end{equation}
The coefficients were obtained from two populations of $2\times10^6$
trajectories.  They predict the shifts at separately evaluated conditions
within $0.605\leq b\leq0.615$ and $0.19\leq M^0\leq0.21$
(Fig.~\ref{fig:reentrance}b), with a largest absolute discrepancy below
$10^{-4}$ in loading.  Each shift is measured relative to the corresponding
population's reference boundary, whose absolute position remains sensitive
to finite sampling.  Stronger correlation raises the loading needed to
leave the mixed-state basin, whereas a larger initial target overlap lowers it.

The upper boundary also reflects loss of basin access in the tested cases.
The target fixed point remains linearly stable across it in all three direct
networks.  At $\alpha=0.05$, DMFT retrieves from $M^0=1$ while the initial
state with $M^0=0.20$ fails, and the tested collective perturbations of
the settled retrieval state decay.  The separating dynamics is less simple here:
edge tracking in two networks gives moving boundary trajectories through
600 updates, with nearby initial states subsequently reaching target and
non-target retrieval.  Their asymptotic boundary dynamics remains unresolved.

\subsection{Cross-talk noise and self-feedback coefficients}

\begin{figure}[htbp]
 \centering
 \includegraphics[width=0.9\textwidth]{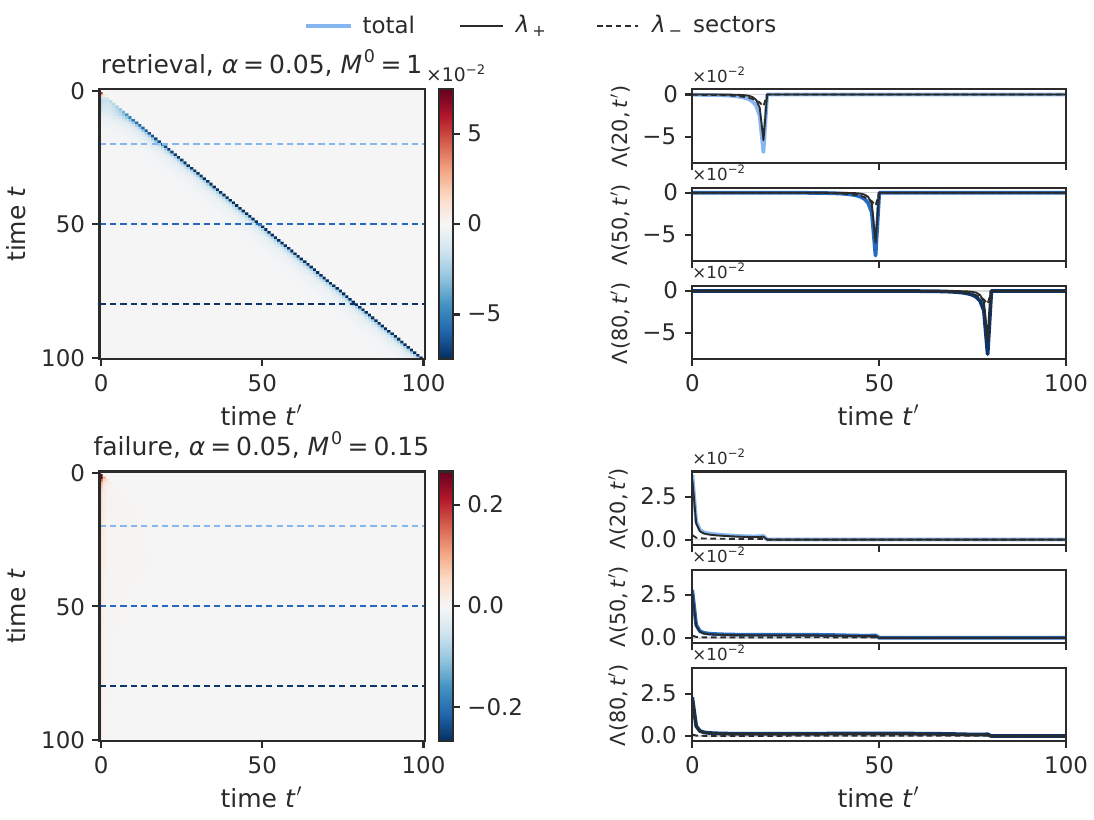}
 \caption{Self-feedback coefficients and contributions from the two
 covariance eigenspaces.  Horizontal guides select the observation times
 shown on the right.  Color and ordinate scales are set separately
 for the two conditions.}
 \label{fig:feedback}
\end{figure}

\begin{figure}[htbp]
 \centering
 \includegraphics[width=0.9\textwidth]{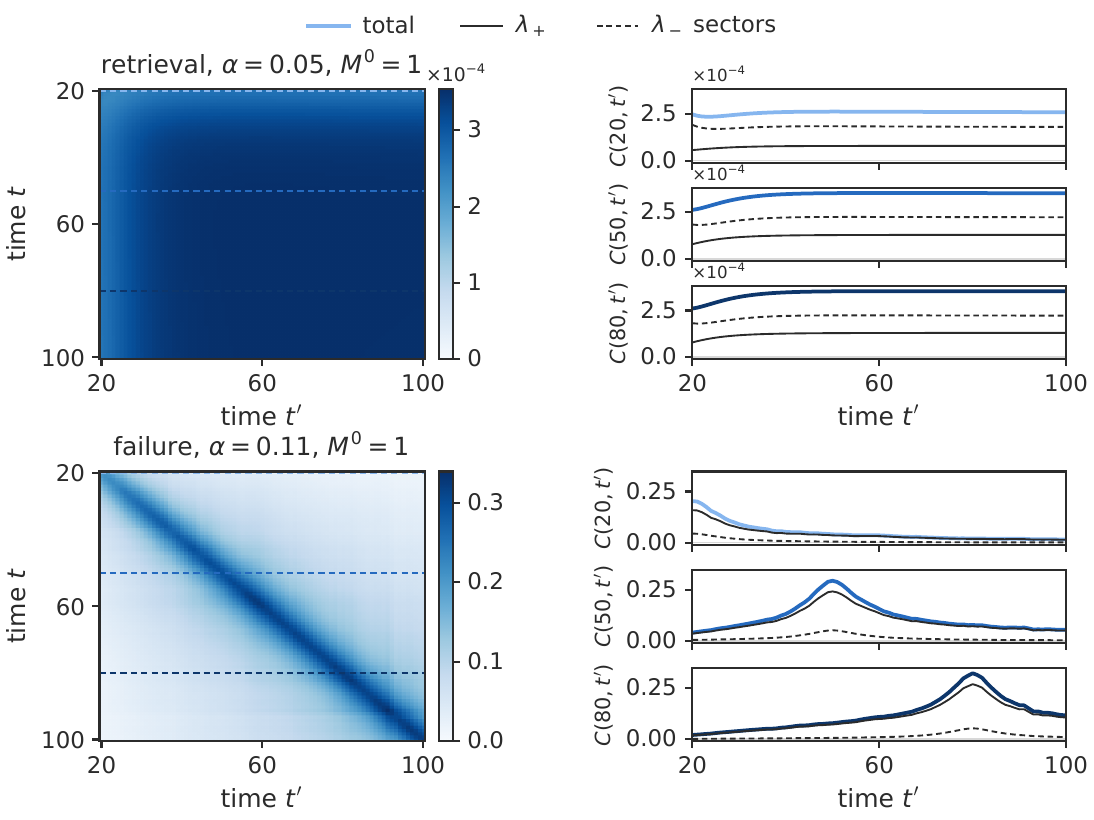}
 \caption{Cross-talk noise covariance for $t,t'\geq20$ and the
 contributions from the covariance eigenspaces.  Horizontal guides select the observation times shown on
 the right.  Color and ordinate scales are set separately for retrieval
 and failure.}
 \label{fig:noise}
\end{figure}

The temporal kernels describe how the uncondensed blocks modify the local
field (Figs.~\ref{fig:feedback} and \ref{fig:noise}).
Each covariance eigenvalue contributes through its own temporal resolvent.

The integrated self-feedback and its stationary limit are
\begin{equation}
 \Lambda(t)=\sum_{t'<t}\Lambda(t,t'),
 \qquad
 \Lambda=\lim_{t\to\infty}\Lambda(t).
 \label{eq:integrated-feedback}
\end{equation}
Let $G(\tau)$ denote the stationary response at lag $\tau$,
$U=\sum_{\tau\geq1}G(\tau)$ the susceptibility (integrated response) and $q$ the late-time
plateau of $Q(t,t')$ over the temporal arguments weighted by the resolvents.
Under the fixed-point assumptions in Appendix~\ref{app:fixed},
\begin{align}
 \Lambda={}&\alpha\left[
 \frac{\lambda_+^2U}{1-\lambda_+U}
 +2\frac{\lambda_-^2U}{1-\lambda_-U}\right],
 \label{eq:result-Lambda-stationary}\\
 \sigma_\phi^2={}&\alpha q\left[
 \frac{\lambda_+^2}{(1-\lambda_+U)^2}
 +2\frac{\lambda_-^2}{(1-\lambda_-U)^2}\right].
 \label{eq:result-sigma-stationary}
\end{align}
These stationary expressions are used below for settled retrieval.

At lag one, the ratio of the $\lambda_+$ contribution to the two
$\lambda_-$ contributions is $\lambda_+^2/(2\lambda_-^2)$.
After temporal summation, the response modifies this ratio by the factor
\begin{equation}
 \frac{1-\lambda_-U}{1-\lambda_+U}.
 \label{eq:feedback-ratio}
\end{equation}
A negative integrated response therefore suppresses the larger covariance
eigenvalue more strongly.  In the retrieval condition, the observed
reduction is consistent with $U\simeq-11.2$, estimated from the late-time
row sum of the $\lambda_+$ resolvent using Eq.~\eqref{eq:B-rowsum}.

The same response suppresses cross-talk through the denominators in
Eq.~\eqref{eq:result-sigma-stationary}.
At the measured $q$, this expression gives a variance about two orders of
magnitude smaller than its value at $U=0$.
This reduction is associated with the Onsager reaction term
\cite{ShiinoFukai1993}.  The residual noise is small but nonzero and
is compatible with a numerically errorless target readout.
The enlarged retrieval scale in Fig.~\ref{fig:noise} makes this residual
covariance visible.

\begin{figure}[htbp]
 \centering
 \includegraphics[width=0.95\textwidth]{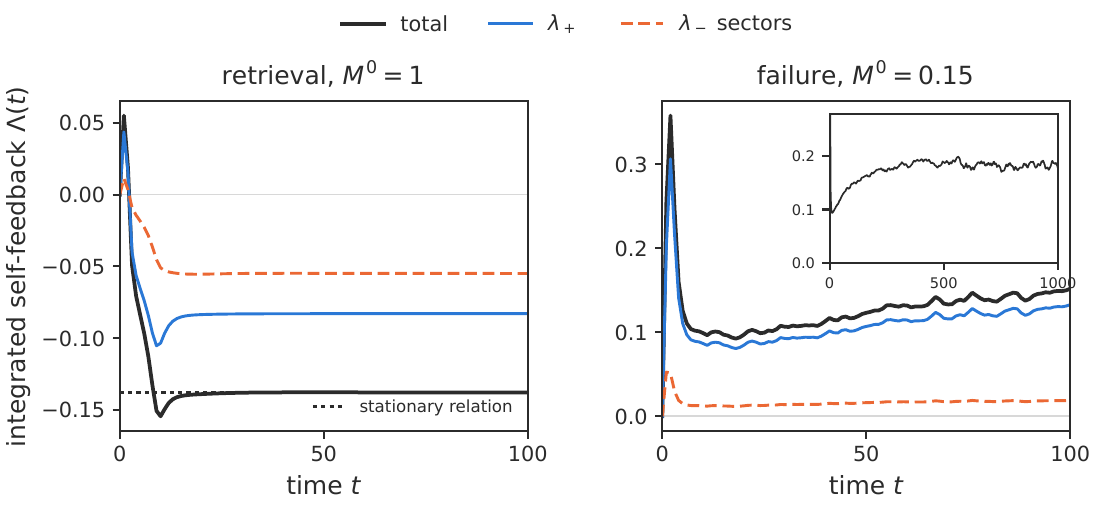}
 \caption{Integrated self-feedback at $\alpha=0.05$ during retrieval and failure.
 The horizontal line in the retrieval panel is
 Eq.~\eqref{eq:result-Lambda-stationary}, with $U$ estimated from the
 late-time resolvent.  The inset extends the failed-retrieval trajectory
 to $T=1000$ using $5\times10^4$ effective trajectories, with the response
 evaluated by Gaussian integration by parts.}
 \label{fig:integrated}
\end{figure}

The failed-retrieval noise covariance decays with time separation, indicating
continued microscopic evolution.  In the failed-retrieval condition at
$(\alpha,M^0)=(0.05,0.15)$, the integrated self-feedback nevertheless
approaches a positive plateau more slowly than in successful retrieval
(Fig.~\ref{fig:integrated}).  Its continued evolution beyond $t=100$ and
eventual saturation agree with the behaviour found for independent patterns
\cite{KabashimaMimura2026}.  Convergence of this integrated coefficient
alone does not imply a neuronal fixed point or justify replacing the
cross-talk by a static Gaussian field.

\subsection{Fixed-point relation and branch selection}

For a stationary output, the effective process gives
\begin{equation}
 x=f(y+\Lambda x),
 \qquad
 y=\bm\xi^{\mathsf T}\bm m+\phi.
 \label{eq:branch}
\end{equation}
If the selected output is single-valued at fixed $y$, it defines an effective
transfer function $x=g(y)$.  The overlap, correlation and susceptibility
averages are then those of Appendix~\ref{app:fixed}.
A unique root of Eq.~\eqref{eq:branch} determines $g$ directly.
When several roots coexist, their selection requires the preceding dynamics;
a single function $g(y)$ suffices only if the histories at the same $y$
select the same output.

\begin{figure}[htbp]
 \centering
 \includegraphics[width=0.95\textwidth]{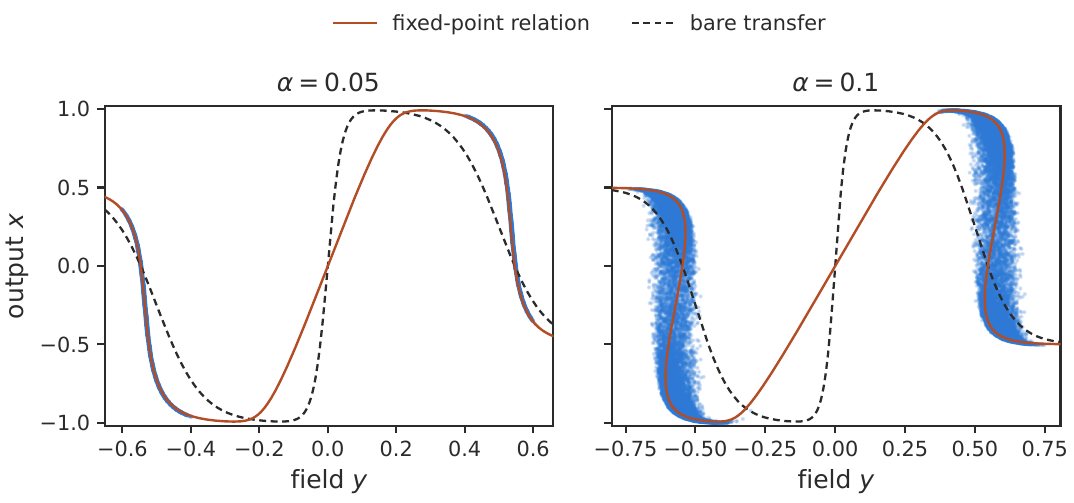}
 \caption{DMFT input-output pairs at $M^0=1$.
 The solid curves give all roots of Eq.~\eqref{eq:branch} with
 $\Lambda$ replaced by $\Lambda(T)$; dashed curves give the bare transfer
 $f(y)$.  Points are the finite-time outputs.}
 \label{fig:branches}
\end{figure}

Figure~\ref{fig:branches} compares the finite-time input-output pairs with
Eq.~\eqref{eq:branch}, using the integrated self-feedback at the observation
time.  At $\alpha=0.05$, the outputs lie near the unique fixed-point root.
At $\alpha=0.10$, the equation admits several roots over part of the sampled
field range.  The outputs have not all settled onto these roots.
The comparison shows why the fixed-point relation alone cannot specify
the observed input-output distribution.

At the lower loading, the target component of $\bm m$ dominates the
non-target components.  The sign of $\xi_1$ therefore separates the local
fields into two groups even though the condensed block has eight binary
configurations.  The multiplicity of fixed-point roots instead arises from
the nonmonotonic transfer and self-feedback, as in the independent-pattern
model \cite{KabashimaMimura2026}.

\subsection{Bistability under sign dynamics}

\begin{figure}[htbp]
 \centering
 \includegraphics[width=\textwidth]{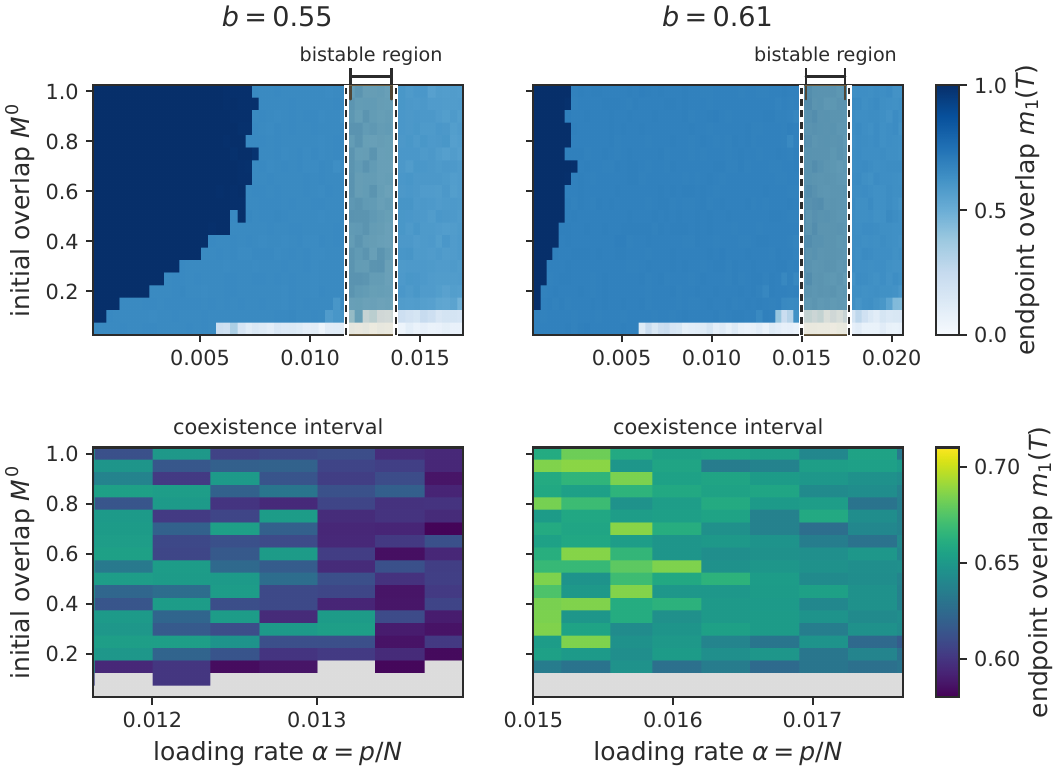}
 \caption{Endpoint overlap under synchronous sign dynamics.
 Outlines in the upper panels mark the coexistence intervals reported in
 Ref.~\cite{ToyaEtAl2000}.  The lower panels enlarge these intervals and
 the overlap range containing the two mixed states.  Values outside this
 overlap range are shown in grey.}
 \label{fig:signbasin}
\end{figure}

We finally set $f(a)=\operatorname{sign}(a)$ and $\gamma=1$ to compare with
Ref.~\cite{ToyaEtAl2000}.  The output and readout overlaps now coincide,
$m_\nu(t)=M_\nu(t)$.  Here $\eta$ denotes the symmetric mixed-state branch
continuing the majority pattern and $\widetilde\eta$ denotes the second
symmetric mixed state.  At finite loading, a state on the $\eta$ branch
need not coincide site by site with the majority pattern.
The response is evaluated by Gaussian integration by parts,
Eq.~\eqref{eq:Grec}, with covariance regularization proportional to the
inverse number of effective trajectories.

The generalized SCSNA predicts coexistence of $\eta$ and $\widetilde\eta$
for $0.01164<\alpha<0.01389$ at $b=0.55$ and
$0.01500<\alpha<0.01765$ at $b=0.61$ \cite{ToyaEtAl2000}.
The same reference demonstrates initial-state dependence and hysteresis in
finite-network simulations.  DMFT determines the corresponding basins in
the thermodynamic limit (Fig.~\ref{fig:signbasin}).
Inside each coexistence interval, the selected overlap changes
nonmonotonically with $M^0$: a larger initial target overlap need not select
the mixed state with the larger final overlap.

\begin{figure}[!htbp]
 \centering
 \includegraphics[width=\textwidth]{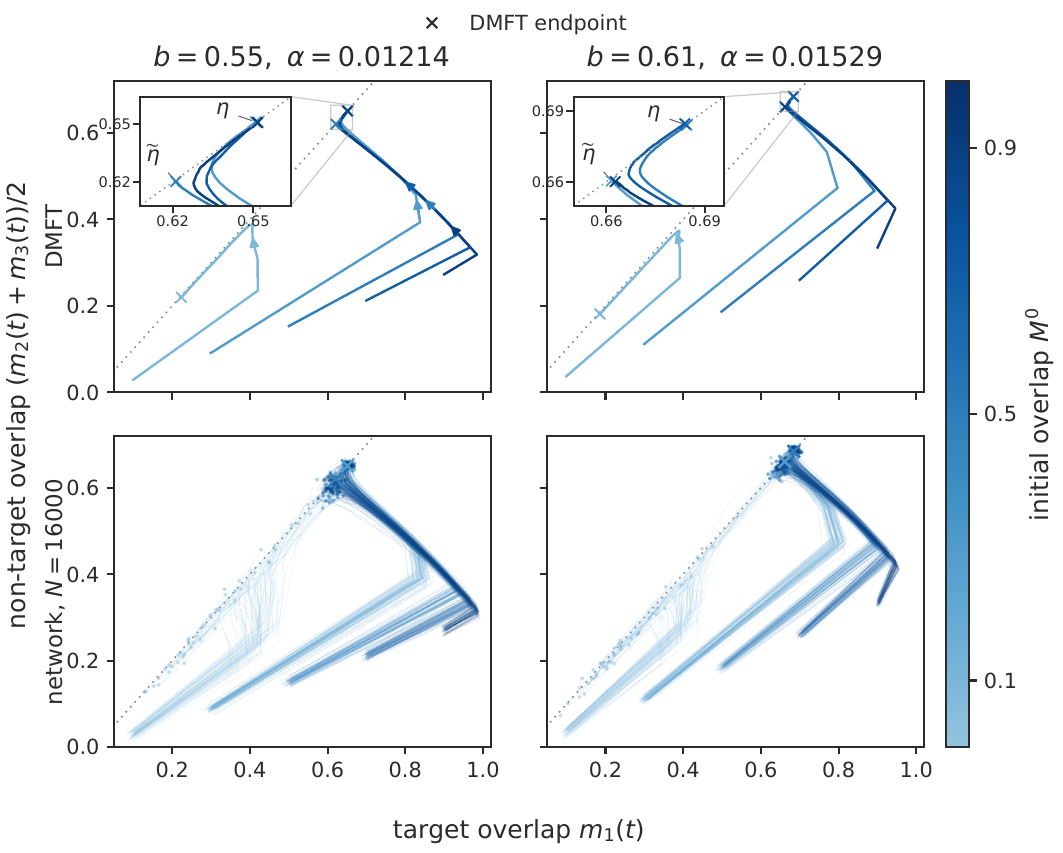}
 \caption{Trajectories under synchronous sign dynamics.
 DMFT uses $10^6$ effective trajectories through $T=300$; direct simulations
 use 64 disorder realizations per initial overlap through $T=100$.
 Crosses mark the DMFT endpoints.  Insets resolve the two symmetric mixed
 states, which lie on the dotted diagonal.}
 \label{fig:signphase}
\end{figure}

The projection onto the target overlap and the mean non-target overlap
$(m_2+m_3)/2$ shows how this selection occurs
(Fig.~\ref{fig:signphase}).  Trajectories with different initial overlaps
can pass through the same neighbourhood before approaching different
points on the symmetric line.  Their different approaches to the two
mixed states account for the nonmonotonic basin dependence.
Direct trajectories follow the same geometry around the DMFT mixed states,
while some reach additional lower-overlap states, consistent with the
spurious states observed in Ref.~\cite{ToyaEtAl2000}.

\FloatBarrier

\section{Summary and discussion}
\label{sec:summary}

We derived the DMFT for an associative memory whose patterns occur in
independent binary blocks of fixed size.  The derivation extends the
independent-pattern generating-functional analysis of
Ref.~\cite{KabashimaMimura2026}.  Diagonalizing the block covariance separates the
uncondensed disorder into covariance eigenmodes.  Each covariance eigenvalue remains
inside the temporal response resolvents that determine the cross-talk noise
covariance and the self-feedback coefficients.  At fixed block size and fixed
finite time, the leading uncondensed contribution depends on the block
distribution through its covariance.  The condensed signal and one-site
averages retain the complete binary distribution.

For the hierarchically correlated patterns, the two
covariance modes have distinct roles in the initial motion.  Under the
linear approximation for weak initial overlaps, the $\lambda_+$ mode amplifies the symmetric block
direction relative to the $\lambda_-$ modes that distinguish one descendant
from the others.  Increasing correlation strengthens this initial preference.
In the full dynamics, the target-retrieval boundaries cross as the correlation
increases.  At the highest correlation examined and at low loading,
the condensed block remains active while descendant discrimination is lost
over the observation time.  Direct finite-network simulations reproduce the
basin geometry and retrieval trajectories over the tested conditions.

Near the re-entrant lower boundary, increasing the loading can restore
retrieval from an imperfect initial state while both the target and symmetric
mixed states remain accessible.  Direct networks reveal stationary saddles
separating the two basins.  DMFT shows an increasing delay before mixed or
target retrieval is selected as the lower boundary is approached.
The response of these trajectories to parameter changes predicts nearby
boundary shifts.  At the upper boundary, target retrieval
remains accessible from other initial states and stable under the tested
perturbations.  The sampled direct boundary trajectories remain time dependent
over the measured interval.  Their asymptotic dynamics remains unresolved.

The response also changes how the covariance modes contribute to cross-talk.
In retrieval, the negative integrated response suppresses the cross-talk noise
variance through the mode-dependent stationary denominators.  The suppression
is strongest in the $\lambda_+$ mode.  Under the stationary assumptions of
Appendix~\ref{app:fixed}, the fixed-point equations reproduce generalized
SCSNA.  When several fixed-point roots coexist, their occupation depends on
the preceding dynamics, as in Ref.~\cite{KabashimaMimura2026}.
A single effective transfer function $g(y)$ is sufficient only when histories
reaching the same field $y$ select the same output.

The generalized SCSNA of Ref.~\cite{ToyaEtAl2000} establishes the stationary mixed
states of the hierarchically correlated ensemble and their coexistence under
sign dynamics.  Retrieval through a mixed state and dependence on the initial
state were already observed in that study.  Here, DMFT determines the basins
of the coexisting mixed states in the thermodynamic limit.  Their selection
depends nonmonotonically on the initial overlap: increasing the initial target
overlap can lead to a mixed state with a smaller final overlap.

Several directions remain for future investigation. A central challenge is to
characterize the dynamical mechanisms governing the boundaries of retrieval
basins directly within DMFT. In particular, the stationary saddles observed in
finite networks near the lower boundary suggest the possibility of an effective
stability analysis of the corresponding macroscopic dynamics, while the
asymptotic behavior of trajectories on the upper boundary remains to be
clarified. Such an analysis could establish a more systematic connection
between the temporal response, basin geometry, and re-entrant retrieval.
Another direction is to examine the extent to which these phenomena depend on
the nonmonotonic transfer function and the specific hierarchical correlation
structure considered here. Applying the present framework to more general
correlated-pattern ensembles, including larger blocks and multiple levels of
hierarchical correlations, may reveal which aspects of the retrieval dynamics
are universal and which depend on the detailed organization of stored memories.

\begin{acknowledgments}
This work was partially supported by the University of Milano-Bicocca through
its Exchange Extra-EU program (AC) and JSPS KAKENHI Grant Nos. 22H05117 and
26K02981, and JST, ARiSE, Japan Grant Number JPMJAR2633 (YK).
\end{acknowledgments}

\appendix

\section{Generating-functional derivation}
\label{app:derivation}

We derive the effective single-site process and its kernels,
Eqs.~\eqref{eq:C} and \eqref{eq:Lambda}, following the generating-functional
calculation in Appendix~A of Ref.~\cite{KabashimaMimura2026}.

\subsection{Constraints and disorder average}

Throughout this appendix, the abbreviated initial distribution
$P_0(\bm s^0)\equiv\prod_iP_0(s_i^0\mid\bm\xi_i^1)$ denotes the
site-factorized distribution conditional on the condensed block.  Its single-site
factor is given by Eq.~\eqref{eq:initial}.  Substitution of
Eq.~\eqref{eq:Fourier} and the field decomposition
Eq.~\eqref{eq:fieldsplit} into Eq.~\eqref{eq:Z} gives
\begin{align}
 Z={}&\sum_{\bm s^0}P_0(\bm s^0)
 \int D\bm a\,D\bm h
 \exp\left[-i\sum_i\sum_{t=0}^{T}h_i(t)
 \{a_i(t+1)-a_i(t)\right.\nonumber\\
 &\left.\quad-\gamma[-a_i(t)+\bm\xi_i^1\!\cdot\bm m(t)
 -s\alpha_Nx_i(t)+\theta_i(t)]\}
 +i\sum_i\sum_{t=0}^{T}\psi_i(t)x_i(t)\right]\nonumber\\
 &\times\prod_{\mu=2}^{p}
 \exp\left[\gamma\sum_{t=0}^{T}\sum_{\nu=1}^{s}
 v_{\mu\nu}(t)u_{\mu\nu}(t)\right],
 \label{eq:A-Zsub}
\end{align}
where $\alpha_N=p/N$ and
\begin{equation}
 D\bm a=\prod_{i=1}^{N}\prod_{t=1}^{T+1}da_i(t),
 \qquad
 D\bm h=\prod_{i=1}^{N}\prod_{t=0}^{T}
 \frac{dh_i(t)}{2\pi}.
 \label{eq:A-measures}
\end{equation}
Here $a_i(0)=0$, $x_i(0)=s_i^0$ and $x_i(t)=f[a_i(t)]$ for
$1\leq t\leq T$.  The final transition introduces $a_i(T+1)$ without an
observable source.  The condensed overlaps at update times $0\leq t\leq T$
are imposed with
\begin{equation}
 1=\frac{N}{2\pi}\int_{-\infty}^{\infty}dm_\nu(t)
 \int_{-\infty}^{\infty}d\widehat m_\nu(t)
 \exp\left\{iN\widehat m_\nu(t)
 \left[m_\nu(t)-\frac1N\sum_i\xi_i^{1,\nu}x_i(t)\right]\right\}.
 \label{eq:A-mconstraint}
\end{equation}
The $d=T+1$ component vectors $\bm x_i$ and $\bm h_i$ collect
$x_i(t)$ and $h_i(t)$, respectively.  We absorb the factors of $i$ into
the conjugate matrices and include the normalization constants in the
integration measures.  The symmetric matrices $Q$, $R$ and their conjugates
are integrated over $0\leq t\leq t'\leq T$.  The half traces below fix the
normalization of the conjugate entries:
\begin{align}
 1={}&\int DQ\,D\widehat Q
 \exp\left\{-\frac{N}{2}\operatorname{Tr}\left[
 \widehat Q\left(Q-\frac1N\sum_i\bm x_i\bm x_i^{\mathsf T}
 \right)\right]\right\},
 \label{eq:A-Qconstraint}\\
 1={}&\int DR\,D\widehat R
 \exp\left\{-\frac{N}{2}\operatorname{Tr}\left[
 \widehat R\left(R-\frac1N\sum_i
 (i\bm h_i)(i\bm h_i)^{\mathsf T}\right)\right]\right\}.
 \label{eq:A-Rconstraint}
\end{align}
The nonsymmetric matrix $S$ and its conjugate are integrated over all
ordered pairs $0\leq t,t'\leq T$:
\begin{equation}
 1=\int DS\,D\widehat S
 \exp\left\{-N\operatorname{Tr}\left[
 \widehat S\left(S-\frac1N\sum_i
 \bm x_i(i\bm h_i)^{\mathsf T}\right)\right]\right\}.
 \label{eq:A-Sconstraint}
\end{equation}

For one uncondensed block, the amplitudes in Eqs.~\eqref{eq:u} and
\eqref{eq:v} have zero mean and the second cumulants below, with
$\widetilde{\bm\xi}$ denoting all stored patterns:
\begin{align}
 \mathbb{E}_{\widetilde{\bm\xi}}[u_{\mu\nu}(t)u_{\mu\nu'}(t')]
 &=B_{\nu\nu'}Q(t,t'),\nonumber\\
 \mathbb{E}_{\widetilde{\bm\xi}}[v_{\mu\nu}(t)v_{\mu\nu'}(t')]
 &=B_{\nu\nu'}R(t,t'),
 \label{eq:A-amplitudecov}\\
 \mathbb{E}_{\widetilde{\bm\xi}}[u_{\mu\nu}(t)v_{\mu\nu'}(t')]
 &=B_{\nu\nu'}S(t,t').\nonumber
\end{align}
These relations hold with the paths and real Fourier variables fixed.  Let
\begin{equation}
 \Sigma=\begin{pmatrix}Q&S\\S^{\mathsf T}&R\end{pmatrix}.
 \label{eq:A-Sigma}
\end{equation}
The second-cumulant matrix of the complete amplitude vector is
$B\otimes\Sigma$, with the factors of $i$ inherited from the definition of
$v$ in Eq.~\eqref{eq:v}.

At fixed block size $s$ and fixed finite time horizon $T$, an amplitude
vector is a finite-dimensional normalized sum of independent site
contributions.  Its cumulant of order $r$ is
$O(N^{1-r/2})$.  The cumulant expansion gives the Gaussian contribution at
leading order.  Higher cumulants produce subextensive corrections after
summing over the $p-1$ uncondensed blocks.

The leading extensive disorder average can therefore be evaluated from the
second cumulants in Eq.~\eqref{eq:A-amplitudecov}.  We diagonalize $B$ as in
Eq.~\eqref{eq:W} and rotate
\begin{equation}
 \widetilde{\bm u}_\mu(t)=W\bm u_\mu(t),
 \qquad
 \widetilde{\bm v}_\mu(t)=W\bm v_\mu(t).
 \label{eq:A-rotation}
\end{equation}
The Jacobian has absolute value one because $W$ is orthogonal.  Moreover,
\begin{equation}
 \bm v_\mu(t)^{\mathsf T}\bm u_\mu(t)
 =\widetilde{\bm v}_\mu(t)^{\mathsf T}
  \widetilde{\bm u}_\mu(t).
 \label{eq:A-bilinear}
\end{equation}
The second-cumulant matrix of covariance eigenmode $k$ is
$\lambda_k\Sigma$.

For one nonzero eigenvalue $\lambda$, dividing the amplitudes by
$\sqrt{\lambda}$ gives the independent-pattern Gaussian integral with
coupling $\gamma\lambda$.  Denote the rescaled time vectors by $\bm u$ and
$\bm v$.  The vectors $\bm x$ and $\bm y$ in the following integral are
auxiliary $d$-component Fourier variables:
\begin{align}
 \mathcal I_\lambda
 ={}&\frac1{(2\pi)^{2d}}
 \int d\bm x\,d\bm y\,d\bm u\,d\bm v
 \exp\left[-\frac12
 (\bm x^{\mathsf T},\bm y^{\mathsf T})
 \Sigma\binom{\bm x}{\bm y}\right.\nonumber\\
 &\left.\qquad+i\bm x^{\mathsf T}\bm u+i\bm y^{\mathsf T}\bm v
 +\gamma\lambda\bm v^{\mathsf T}\bm u\right]
 \label{eq:A-gauss1}\\
 ={}&\frac1{(2\pi)^d}
 \int d\bm x\,d\bm u
 \exp\left[-\frac12\bm x^{\mathsf T}Q\bm x
 +i\bm x^{\mathsf T}(I-\gamma\lambda S)\bm u
 +\frac{\gamma^2\lambda^2}{2}\bm u^{\mathsf T}R\bm u\right]
 \label{eq:A-gauss2}\\
 ={}&\frac1{(2\pi)^{d/2}(\det Q)^{1/2}}
 \int d\bm u\,
 \exp\left[-\frac12\bm u^{\mathsf T}
 \left\{(I-\gamma\lambda S)^{\mathsf T}Q^{-1}
 (I-\gamma\lambda S)-\gamma^2\lambda^2R\right\}\bm u\right]
 \label{eq:A-gauss3}\\
 ={}&\left\{\det Q\,
 \det\left[(I-\gamma\lambda S)^{\mathsf T}Q^{-1}
 (I-\gamma\lambda S)-\gamma^2\lambda^2R\right]
 \right\}^{-1/2}.
 \label{eq:A-gauss4}
\end{align}
The integrations over $\bm v$ and $\bm y$ give the second line.  Integration
over $\bm x$ gives the third line and the final Gaussian integral gives the
determinant in Eq.~\eqref{eq:A-gauss4}.  A zero eigenvalue contributes one.
For singular $Q$, we use $Q+\epsilon I$ and take $\epsilon\downarrow0$ after
evaluating the saddle.

Multiplying this determinant over the covariance eigenmodes and the
uncondensed blocks gives
\begin{align}
 &\prod_{k=1}^{s}
 \left\{\det Q\,
 \det\left[(I-\gamma\lambda_kS)^{\mathsf T}Q^{-1}
 (I-\gamma\lambda_kS)-\gamma^2\lambda_k^2R\right]
 \right\}^{-(p-1)/2}.
 \label{eq:A-pminus1}
\end{align}
Since $(p-1)/N\to\alpha$, its logarithm gives
Eq.~\eqref{eq:detproduct} at order $N$.

\subsection{Extensive action and saddle point}

We now combine the determinant with the constraint integrals and average
over the stored patterns.  Collecting all terms of order $N$ gives
\begin{equation}
 \mathbb{E}_{\widetilde{\bm\xi}}[Z[\bm\psi,\bm\theta]]
 =\int D\Omega\,\exp\left\{N\Psi_{\rm mac}[\Omega]
 +\sum_{i=1}^{N}\log Z_{1,i}[\Omega;\bm\psi_i,\bm\theta_i]
 +o(N)\right\},
 \label{eq:A-action}
\end{equation}
where $D\Omega$ contains $\bm m,Q,R,S$ and their conjugates.  The macroscopic
part is
\begin{align}
 \Psi_{\rm mac}={}&i\sum_{\nu=1}^{s}\sum_{t=0}^{T}
 \widehat m_\nu(t)m_\nu(t)
 -\frac12\operatorname{Tr}(\widehat Q Q+\widehat R R)
 -\operatorname{Tr}(\widehat S S)\nonumber\\
 &-\frac{\alpha}{2}\sum_{k=1}^{s}
 \log\det Q
 -\frac{\alpha}{2}\sum_{k=1}^{s}
 \log\det\left[(I-\gamma\lambda_kS)^{\mathsf T}Q^{-1}
 (I-\gamma\lambda_kS)-\gamma^2\lambda_k^2R\right].
 \label{eq:A-Psi}
\end{align}
All traces and determinants in Eq.~\eqref{eq:A-Psi} run over
$0\leq t,t'\leq T$.  For the one-site trace we use
\begin{equation}
 Da=\prod_{t=1}^{T+1}da(t),
 \qquad
 Dh=\prod_{t=0}^{T}\frac{dh(t)}{2\pi}.
 \label{eq:A-one-site-measures}
\end{equation}
The site-dependent trace retains the complete condensed distribution,
\begin{align}
 Z_{1,i}={}&\sum_{\bm\xi}P(\bm\xi)
 \sum_{s^0}P_0(s^0\mid\bm\xi)
 \int Da\,Dh
 \exp\left[-i\sum_{t=0}^{T}h(t)
 \{a(t+1)-a(t)\right.\nonumber\\
 &\left.\quad-\gamma[-a(t)+\bm\xi^{\mathsf T}\bm m(t)
 -s\alpha x(t)+\theta_i(t)]\}
 +i\sum_{t=0}^{T}\psi_i(t)x(t)
 -i\sum_{\nu=1}^{s}\sum_{t=0}^{T}
 \widehat m_\nu(t)\xi_\nu x(t)\right.\nonumber\\
 &\left.\quad+\frac12\sum_{t,t'=0}^{T}
 x(t)\widehat Q(t,t')x(t')
 +\frac12\sum_{t,t'=0}^{T}
 [i h(t)]\widehat R(t,t')[i h(t')]\right.\nonumber\\
 &\left.\quad+\sum_{t,t'=0}^{T}
 [i h(t)]\widehat S(t,t')x(t')\right].
 \label{eq:A-Z1}
\end{align}
At zero sources, all sites have the same representative trace $Z_1$.
Variation with
respect to $\widehat m$ gives Eq.~\eqref{eq:mself}, while variation with
respect to $\bm m$ gives $\widehat{\bm m}=0$ at the normalized saddle.

Normalization determines $R$ and $\widehat Q$.
The identity $Z[\bm0,\bm\theta]=1$ gives
\begin{equation}
 0=\frac{\partial^2 Z}{\partial\theta_i(t)\partial\theta_i(t')}
 =\gamma^2\langle i h_i(t)i h_i(t')\rangle
 \label{eq:A-Rzero}
\end{equation}
for $0\leq t,t'\leq T$ at zero observable source.  Here angular brackets
denote normalized averages in the Fourier representation.  Therefore $R=0$.
The two determinants at this
saddle satisfy
\begin{equation}
 \log\det Q+\log\det\left[
 (I-\gamma\lambda_kS)^{\mathsf T}Q^{-1}
 (I-\gamma\lambda_kS)\right]
 =2\log\det(I-\gamma\lambda_kS).
 \label{eq:A-detnormal}
\end{equation}
The result is independent of $Q$, which gives $\widehat Q=0$.

The remaining conjugate saddle values follow by differentiating the
determinant terms and evaluating at $R=0$:
\begin{align}
 &\left.\delta_R\log\det\left[
 (I-\gamma\lambda_kS)^{\mathsf T}Q^{-1}
 (I-\gamma\lambda_kS)-\gamma^2\lambda_k^2R\right]\right|_{R=0}
 \nonumber\\
 &\quad=-\gamma^2\lambda_k^2
 \operatorname{Tr}\left[
 (I-\gamma\lambda_kS)^{-1}Q
 (I-\gamma\lambda_kS^{\mathsf T})^{-1}\delta R\right],
 \label{eq:A-varR}\\
 \delta_S[-\log\det(I-\gamma\lambda_kS)]
 ={}&\gamma\lambda_k
 \operatorname{Tr}[(I-\gamma\lambda_kS)^{-1}\delta S].
 \label{eq:A-varS}
\end{align}
The $S$ variation of the normalized action is therefore
\begin{equation}
 \delta_S\Psi_{\rm mac}
 =\operatorname{Tr}\left\{\left[-\widehat S
 +\alpha\gamma\sum_{k=1}^{s}\lambda_k
 (I-\gamma\lambda_kS)^{-1}\right]\delta S\right\}.
 \label{eq:A-Ssaddle}
\end{equation}
Stationarity under the $R$ and $S$ variations gives
\begin{align}
 \widehat R={}&\alpha\gamma^2\sum_{k=1}^{s}\lambda_k^2
 (I-\gamma\lambda_kS)^{-1}Q
 (I-\gamma\lambda_kS^{\mathsf T})^{-1},
 \label{eq:Rhat}\\
 \widehat S={}&\alpha\gamma\sum_{k=1}^{s}\lambda_k
 (I-\gamma\lambda_kS)^{-1}.
 \label{eq:Shat}
\end{align}
Together with the definitions of $Q$, $R$, $S$ and $\bm m$, these equations
close the saddle point.

\subsection{Gaussian decoupling and response}

Differentiating the single-site Fourier trace in Eq.~\eqref{eq:A-Z1} with
respect to the external field gives
\begin{equation}
 G(t,t')=\left.\frac{\partial\langle x(t)\rangle}
 {\partial\theta(t')}\right|_{\bm\theta=0}
 =\gamma\langle x(t)ih(t')\rangle
 =\gamma S(t,t').
 \label{eq:A-sourceGS}
\end{equation}
Here the brackets denote normalized Fourier averages in $Z_1$.  The
macroscopic saddle values are held fixed in this local derivative,
corresponding to fixed $\bm m$, $C$ and $\Lambda$ in
Eq.~\eqref{eq:Gself}.  Since the external field at $t'$ first affects the
output at $t'+1$, $G$ is strictly lower triangular.

The quadratic Fourier-field term can now be written as a Gaussian average,
\begin{equation}
 \exp\left[\frac12(i\bm h)^{\mathsf T}\widehat R(i\bm h)\right]
 =\mathbb{E}_{\bm\phi}\exp\left[i\gamma\bm h^{\mathsf T}\bm\phi\right],
 \qquad C=\gamma^{-2}\widehat R,
 \label{eq:A-decoupling}
\end{equation}
where $\bm\phi$ is a zero-mean Gaussian process with covariance $C$,
independent of the condensed block and the initial state.  Substituting
$G=\gamma S$ into Eq.~\eqref{eq:Rhat} gives Eq.~\eqref{eq:C}.
The self-interaction matrix in Eq.~\eqref{eq:A-Z1} becomes
\begin{align}
 -\alpha sI+\gamma^{-1}\widehat S
 &=-\alpha sI+\alpha\sum_k\lambda_k
 (I-\lambda_kG)^{-1}\nonumber\\
 &=\alpha\left(\sum_k\lambda_k-s\right)I
 +\alpha\sum_k\lambda_k^2G(I-\lambda_kG)^{-1}.
 \label{eq:A-cancel1}
\end{align}
The instantaneous term vanishes because $\operatorname{Tr}B=s$.  The
remaining term is the self-feedback matrix $\Lambda$ of
Eq.~\eqref{eq:Lambda}, with $\Lambda(t,t)=0$ by causality.

Integration over the Fourier variables therefore gives
\begin{align}
 Z_{\rm eff}[\psi,\theta]={}&\sum_{\bm\xi}P(\bm\xi)
 \sum_{s^0}P_0(s^0\mid\bm\xi)
 \mathbb{E}_{\bm\phi}\Biggl[
 \int Da\prod_{t=0}^{T}\delta\Biggl\{a(t+1)-a(t)\nonumber\\
 &\quad-\gamma\left[-a(t)+\bm\xi^{\mathsf T}\bm m(t)
 +\theta(t)+\phi(t)+\sum_{t'=0}^{t-1}\Lambda(t,t')x(t')\right]\Biggr\}
 \nonumber\\
 &\quad\times\exp\left[i\sum_{t=0}^{T}\psi(t)x(t)\right]\Biggr].
 \label{eq:A-effectiveZ}
\end{align}
The delta functions impose Eq.~\eqref{eq:Langevin}.  At $\psi=0$, each
transition integrates to one, giving $Z_{\rm eff}[0,\theta]=1$.
The saddle identities for $Q$ and $S$ become the self-consistency equations
\eqref{eq:Qself} and \eqref{eq:Gself} for this process.

At fixed $\bm m$, $C$ and $\Lambda$, the external field and the Gaussian
noise enter the dynamics in the same way.  Gaussian integration by parts
therefore relates the correlation between output and noise to the local response:
\begin{align}
 \mathbb{E}_{\bm\xi,s^0,\bm\phi}[x(t)\phi(u)]
 &=\sum_{t'<t}C(t',u)
 \left.\frac{\partial\mathbb{E}_{\bm\xi,s^0,\bm\phi}[x(t)]}
 {\partial\theta(t')}\right|_{\bm\theta=0}\nonumber\\
 &=\sum_{t'<t}G(t,t')C(t',u).
 \label{eq:A-Novikov}
\end{align}
For the principal submatrix $C_{t-1}$ on times $0,\ldots,t-1$, inversion
gives Eq.~\eqref{eq:Grec} when $C_{t-1}$ is nonsingular.  If it is singular,
differentiation with respect to the local source remains the exact
definition of $G$.  A pseudoinverse recovers only the component in the
covariance range.  Finally, the binary initial state gives $Q(0,0)=1$ and
hence the initial noise covariance in Eq.~\eqref{eq:C00}.

\section{Fixed-point conditions and generalized SCSNA}
\label{app:fixed}

We consider solutions of the effective dynamics that converge to a stable
fixed point.  We assume that the late-time response depends on the time lag
$\tau=t-t'$ and has a finite susceptibility (integrated response)
\begin{equation}
 U=\sum_{\tau=1}^{\infty}G(\tau).
 \label{eq:B-U}
\end{equation}
The transient contributions to the time sums are assumed to vanish, and the
row sums of $(I-\lambda_kG)^{-1}$ to converge.  These assumptions allow the
late-time limit to be taken inside the response convolutions below.

The finite-time identity
\begin{equation}
 (I-\lambda_kG)^{-1}
 =I+\lambda_kG(I-\lambda_kG)^{-1}
 \label{eq:B-identity}
\end{equation}
gives, on summing row $t$,
\begin{equation}
 \sum_{u=0}^{t}[(I-\lambda_kG)^{-1}]_{tu}
 =1+\lambda_k\sum_{v<t}G(t,v)
 \sum_{u=0}^{v}[(I-\lambda_kG)^{-1}]_{vu}.
 \label{eq:B-row-equation}
\end{equation}
At late times, the inner row sum on the right approaches the same limit as
the row sum on the left.  Using Eq.~\eqref{eq:B-U} and solving the resulting
scalar equation gives
\begin{equation}
 \lim_{t\to\infty}\sum_{u=0}^{t}[(I-\lambda_kG)^{-1}]_{tu}
 =\frac1{1-\lambda_kU}.
 \label{eq:B-rowsum}
\end{equation}
We take $\lambda_kU<1$ for every nonzero covariance eigenvalue.  The row-sum
argument applies also to large negative $U$ whenever the assumed limits
exist.

For the noise covariance, we further assume that $Q(u,v)$ approaches the
stationary second moment $q$ over the times contributing to both resolvent
sums in Eq.~\eqref{eq:C}.  The two time sums then factorize:
\begin{equation}
 \sigma_\phi^2=\lim_{t,t'\to\infty}C(t,t')
 =\alpha q\sum_{k=1}^{s}\lambda_k^2
 \left\{\lim_{t\to\infty}\sum_{u=0}^{t}
 [(I-\lambda_kG)^{-1}]_{tu}\right\}^{2}.
 \label{eq:B-Ccontraction}
\end{equation}
Substitution of Eq.~\eqref{eq:B-rowsum} gives
\begin{equation}
 \boxed{
 \sigma_\phi^2=\alpha q\sum_{k=1}^{s}
 \frac{\lambda_k^2}{(1-\lambda_kU)^2}.}
 \label{eq:B-sigma}
\end{equation}
The integrated self-feedback coefficient and its stationary limit are
\begin{equation}
 \Lambda(t)=\sum_{t'<t}\Lambda(t,t'),
 \qquad
 \Lambda=\lim_{t\to\infty}\Lambda(t).
 \label{eq:B-integrated-feedback}
\end{equation}
Summing Eq.~\eqref{eq:Lambda} over the earlier time gives
\begin{equation}
 \boxed{\begin{aligned}
 \Lambda&=\alpha\sum_{k=1}^{s}\lambda_k
 \left(\frac1{1-\lambda_kU}-1\right)\\
 &=\alpha\sum_{k=1}^{s}
 \frac{\lambda_k^2U}{1-\lambda_kU}.
 \end{aligned}}
 \label{eq:B-Lambda}
\end{equation}

The limiting cross-talk noise is a static Gaussian field
$\phi=\sigma_\phi z$, where $z$ is a standard Gaussian variable.  The
stationary form of Eq.~\eqref{eq:Langevin} is
\begin{equation}
 a=\bm\xi^{\mathsf T}\bm m+\sigma_\phi z+\Lambda x,
 \qquad x=f(a).
 \label{eq:B-field}
\end{equation}
If the dynamics selects a single output at each field
$y=\bm\xi^{\mathsf T}\bm m+\sigma_\phi z$, it defines an effective transfer
function $g$ satisfying
\begin{align}
 x&=g(\bm\xi^{\mathsf T}\bm m+\sigma_\phi z),
 \label{eq:B-g}\\
 g(y)&=f[y+\Lambda g(y)].
 \label{eq:B-geq}
\end{align}
Using the Gaussian measure $Dz=(2\pi)^{-1/2}e^{-z^2/2}dz$, the
stationary averages are \cite{KabashimaMimura2026}
\begin{align}
 m_\nu&=\sum_{\bm\xi}P(\bm\xi)\int Dz\,
 \xi_\nu g(\bm\xi^{\mathsf T}\bm m+\sigma_\phi z),
 \label{eq:B-m}\\
 q&=\sum_{\bm\xi}P(\bm\xi)\int Dz\,
 g^2(\bm\xi^{\mathsf T}\bm m+\sigma_\phi z).
 \label{eq:B-q}
\end{align}

To obtain $U$, apply a constant external field $\theta$ at late times,
with $\bm m$, $\sigma_\phi$ and $\Lambda$ held fixed.  Assume that its effect
on the selected output is the translation $g(y)\mapsto g(y+\theta)$.
The integrated response is then
\begin{align}
 U&=\left.\frac{\partial}{\partial\theta}
 \sum_{\bm\xi}P(\bm\xi)\int Dz\,
 g(\bm\xi^{\mathsf T}\bm m+\sigma_\phi z+\theta)
 \right|_{\theta=0}
 \label{eq:B-Uresponse}\\
 &=\frac1{\sigma_\phi}\sum_{\bm\xi}P(\bm\xi)\int Dz\,
 z g(\bm\xi^{\mathsf T}\bm m+\sigma_\phi z).
 \label{eq:B-susceptibility}
\end{align}
For $\sigma_\phi>0$, the second line follows by shifting the integration
variable by $\theta/\sigma_\phi$ and differentiating the Gaussian density.
This form also applies to discontinuous effective transfer functions, such
as the sign function.  For $\sigma_\phi=0$, the response is defined by
Eq.~\eqref{eq:B-Uresponse}.

A unique root of Eq.~\eqref{eq:B-geq} determines $g$ directly.  When several
roots coexist, the initial state and transient dynamics determine which
branch is selected \cite{KabashimaMimura2026}.  If histories at the same
$y$ select different outputs, Eqs.~\eqref{eq:B-m} and \eqref{eq:B-q} require
separate averages of $x$ and $x^2$ over these histories.  The susceptibility
must likewise include the response of the branch selection to the external
field.

For the hierarchically correlated patterns, Eqs.~\eqref{eq:B-sigma} and
\eqref{eq:B-Lambda} become
\begin{align}
 \sigma_\phi^2={}&\alpha q\left[
 \frac{\lambda_+^2}{(1-\lambda_+U)^2}
 +(s-1)\frac{\lambda_-^2}{(1-\lambda_-U)^2}\right],
 \label{eq:B-sigmapm}\\
 \Lambda={}&\alpha\left[
 \frac{\lambda_+^2U}{1-\lambda_+U}
 +(s-1)\frac{\lambda_-^2U}{1-\lambda_-U}\right].
 \label{eq:B-Lambdapm}
\end{align}
These expressions reproduce the generalized SCSNA for Model 1 of Toya
\textit{et al.}\ \cite{ToyaEtAl2000}, whose loading $\alpha$ also counts
blocks.  Their cross-talk parameter $r$ and Onsager reaction term $\Gamma$
are identified by
\begin{equation}
 \alpha_{\mathrm{Toya}}=\alpha,
 \qquad
 r=\frac{\sigma_\phi^2}{\alpha},
 \qquad
 \Gamma=\Lambda.
 \label{eq:B-map}
\end{equation}
Their effective output $Y$ therefore obeys
\begin{equation}
 Y=f\left(\sum_{\nu=1}^{s}\xi_\nu m_\nu
 +\Gamma Y+\sqrt{\alpha r}\,z\right).
 \label{eq:B-Y}
\end{equation}
Together with Eqs.~\eqref{eq:B-m}, \eqref{eq:B-q} and
\eqref{eq:B-susceptibility}, this gives their overlap, output-correlation
and susceptibility equations.  In the generic formulation of their
Sec.~III, the loading is $\widehat\alpha=s\alpha$ and the corresponding
noise parameter is $r/s$.

For their sign-transfer model, the Maxwell rule selects
$g(y)=\operatorname{sign}(y)$.  With $q=1$, the Gaussian averages then give
their Eqs.~(38)--(41) for the equilibrium states.

\bibliography{references}

\end{document}